\documentclass{aa}

\usepackage[english]{babel}
\usepackage[T1]{fontenc}
\usepackage[varg]{txfonts}
\usepackage{mathptmx}
\usepackage{amsmath}
\usepackage{amssymb}
\usepackage{amstext}
\usepackage{amsfonts}
\usepackage{mathrsfs}
\usepackage{graphicx}
\usepackage{float}
\usepackage{hyperref}
\usepackage{bm}
\bibpunct{(}{)}{;}{a}{}{,}
\newcommand{\mockalph}[1]{}

\newcommand{\g}{\textrm{g}}
\newcommand{\mm}{\textrm{mm}}
\newcommand{\cm}{\textrm{cm}}
\newcommand{\m}{\textrm{m}}
\newcommand{\au}{\textrm{au}}
\newcommand{\s}{\textrm{s}}
\newcommand{\yr}{\textrm{yr}}
\newcommand{\kyr}{\textrm{kyr}}
\newcommand{\K}{\textrm{K}}
\newcommand{\del}{\partial}

\begin{document}

\title{On the coexistence of the streaming instability and the vertical shear instability in protoplanetary disks}
\author{Urs Sch{\"a}fer{\inst{\ref{Hamburg},\ref{Lund}}}
\and Anders Johansen\inst{\ref{Lund}}
\and Robi Banerjee\inst{\ref{Hamburg}}}
\institute{Hamburg Observatory, University of Hamburg, Gojenbergsweg 112, 21029 Hamburg, Germany,\\\email{urs.schaefer@hs.uni-hamburg.de}\label{Hamburg}
\and Lund Observatory, Department of Astronomy and Theoretical Physics, Lund University, Box 43, 22100 Lund, Sweden\label{Lund}}
\date{}
\abstract{The streaming instability is a leading candidate mechanism to explain the formation of planetesimals. Yet, the role of this instability in the driving of turbulence in protoplanetary disks, given its fundamental nature as a linear hydrodynamical instability, has so far not been investigated in detail. We study the turbulence that is induced by the streaming instability as well as its interaction with the vertical shear instability. For this purpose, we employ the FLASH Code to conduct two-dimensional axisymmetric global disk simulations spanning radii from~$1~\au$ to~$100~\au$, including the mutual drag between gas and dust as well as the radial and vertical stellar gravity. If the streaming instability and the vertical shear instability start their growth at the same time, we find the turbulence in the dust mid-plane layer to be primarily driven by the streaming instability. It gives rise to vertical gas motions with a Mach number of up to~${\sim}10^{-2}$. The dust scale height is set in a self-regulatory manner to about~$1\%$ of the gas scale height. In contrast, if the vertical shear instability is allowed to saturate before the dust is introduced into our simulations, then it continues to be the main source of the turbulence in the dust layer. The vertical shear instability induces turbulence with a Mach number of~${\sim}10^{-1}$ and thus impedes dust sedimentation. Nonetheless, we find the vertical shear instability and the streaming instability in combination to lead to radial dust concentration in long-lived accumulations which are significantly denser than those formed by the streaming instability alone. Thus, the vertical shear instability may promote planetesimal formation by creating weak overdensities that act as seeds for the streaming instability.}
\keywords{hydrodynamics -- instabilities -- turbulence -- methods: numerical -- planets and satellites: formation -- protoplanetary disks}
\titlerunning{Coexistence of streaming instability and vertical shear instability}
\authorrunning{Sch{\"a}fer et al.}
\maketitle

\section{Introduction}
\label{sect:introduction}
\subsection{Observations}
Turbulence crucially influences various stages of the formation of planets: (1) the vertical settling of dust grains to a mid-plane layer, whose thickness is determined by the equilibrium between sedimentation and turbulent diffusion \citep{Dubrulle1995, Johansen2005, Fromang2006, Youdin2007b}; (2) collisional grain growth \citep{Ormel2007, Birnstiel2010} and (3) planetesimal formation owing to passive concentration of grains \citep{Barge1995, Johansen2007b, Cuzzi2008}; as well as (4) planetary migration \citep{Nelson2004, Oishi2007, Yang2009, Yang2012, Baruteau2011}.

It is challenging, though, to observationally constrain the strength of the turbulence in the gas and the dust in protoplanetary disks, whose motions are coupled via drag. This is particularly because these turbulent motions are weaker than both the thermal gas motions and the orbital motions of gas and dust \citep{Flaherty2018}.

Only recently, high-resolution ALMA observations of protoplanetary disks have permitted a number of authors to assess the turbulent strength in these disks: \citet{Flaherty2015, Flaherty2017, Flaherty2018} derive upper limits of the strength of the turbulent vertical gas motions in the disks surrounding HD 163296 and TW Hya from the non-thermal broadening of molecular emission lines. These upper limits correspond to Mach numbers of the order of~$0.01$\footnote{Here and in the following, we report turbulent strength in terms of Mach numbers rather than turbulent~$\alpha$-parameters \citep{Shakura1973}. This is to avoid confusion with the~$\alpha$-parameter associated with angular momentum transport. Based on a mixing length approach and assuming that the eddy turn-over time scale is similar to the inverse of the orbital frequency -- a valid assumption for the instabilities that we investigate in this paper, the streaming instability \citep{Youdin2005} and the vertical shear instability \citep{Nelson2013} -- the turbulent~$\alpha$-parameter can be approximated as the square of the Mach number.}. 

\citet{Pinte2016} employ a model of micron- to millimeter-sized grains in the disk around HL Tau to estimate the strength of their vertical turbulent diffusion from the observed dust scale height, which is equal to~${\sim}10\%$ of the gas scale height. They as well find a Mach number of~${\sim}0.01$. Similarly, \citet{Ohashi2019} constrain the dust grain sizes and dust scale height in the disk around HD 163296 using polarization measurements of the dust emission. From a grain size of~${\sim}100~\mu\m$ and a scale height of less than one third of the gas scale height inside the ring that is located at an orbital radius of~$70~\au$ in this disk, they infer an upper limit of the Mach number of the vertical gas velocity of~${\sim}0.01$. On the other hand, from the dust-to-gas scale height ratio of two thirds outside of the ring they estimate a Mach number of the order of~$0.1$.

Furthermore, \citet{Dullemond2018} obtain a lower limit of the strength of the radial turbulent motions from the width of the dust rings that characterize a majority of the observed disks. This limit as well amounts to a Mach number of about~$0.01$ for grains with a size of~$0.2~\mm$, but is proportional to the grain size.

In addition to the uncertainties in the observational determination of the turbulent strength, it remains to theoretically establish which physical processes are the dominant sources of protoplanetary disk turbulence. In large fractions of disks, particularly in the mid-plane, magnetohydrodynamic (MHD) turbulence is suppressed because of non-ideal MHD effects \citep{Gressel2015, Bai2017}. Turbulence in these regions must therefore be driven either from the MHD turbulent disk surface \citep{Oishi2009, Bai2015} or by purely hydrodynamic instabilites.

\subsection{Vertical shear instability}
One of the most promising among the hydrodynamical instabilities is the vertical shear instability \citep{Arlt2004, Nelson2013}, which is similar to the Goldreich-Schubert-Fricke instability in differentially rotating stars \citep{Goldreich1967, Fricke1968}. It arises when the disk rotation profile depends on the height. This can, for instance, be due to baroclinity, i.e., a misalignment between the density and the pressure gradient, resulting from a radial temperature profile. The source of energy of the instability is the free energy associated with the vertical shear \citep{Barker2015}.

The vertical shear instability can overcome both the radial angular momentum gradient, as its modes are characterized by a large radial-to-vertical wavenumber ratio, and the vertical buoyancy if the gas cooling time scale is sufficiently short \citep{Nelson2013, Lin2015}. Analytical and numerical analyses of the linear growth of the instability have found two classes of modes: short-wavelength surface modes with higher growth rates and vertically global body modes with lower growth rates. The former appear at the artificial vertical simulation domain boundary, where the vertical shear is strongest -- although natural transitions in the density can as well give rise to these modes. Their growth rate increases with the vertical shear at the boundary, and thus with vertical domain size \citep{Nelson2013, Barker2015, Lin2015}.

In non-linear simulations, the vertical shear instability grows over at least~\mbox{$\sim30$} orbital periods until it attains at saturated state \citep{Stoll2014, Flock2017}. The Mach numbers of the turbulent vertical motions in this state are of the order of~$10^{-2}$ to~$10^{-1}$ \citep{Flock2017}. This turbulent strength is higher if the radial temperature gradient is steeper \citep{Nelson2013, Lin2019}. Perturbations associated with the surface modes appear first and grow towards the disk mid-plane. The later emerging body modes are characterized by perturbations that evolve from an odd symmetry with respect to the mid-plane into an even symmetry. Hence, the instability saturates last in this plane \citep{Nelson2013, Stoll2014}. 

The turbulence that is induced by the vertical shear instability entails angular momentum being transported radially outwards and vertically away from the mid-plane. Since the latter eliminates the vertical shear, external heating (of locally non-isothermal disks) is necessary to sustain the instability \citep{Stoll2014}. In numerical models including dust and the drag exerted on it by the gas, the turbulence further gives rise to vertical dust motions, and radial motions leading to accumulation in overdensities of up to five times the initial dust density \citep{Stoll2016, Flock2017}. 

However, \citet{Lin2017} and \citet{Lin2019} show that, if the drag back-reaction of the dust onto the gas is taken into account as well, the dust, which sediments to the mid-plane, introduces an effective vertical buoyancy -- it ``weighs down'' the gas -- that can quench the vertical shear instability. If they are tightly coupled, gas and dust can be described as a single fluid, with a density equal to the sum of the gas and the dust density, but a pressure that is solely due to the gas. We assume that the pure gas is locally isothermal, i.e., its cooling time scale is infinitely short. Then, the equation of state of the mixture is given by~\mbox{$P=c_{\rm s}^2\rho_{\rm g}=c_{\rm s}^2(1-\rho_{\rm d}/\rho_{\rm tot})\rho_{\rm tot}$}, where~$P$ is the gas pressure,~$c_{\rm s}$ the sound speed, and~$\rho_{\rm g}$,~$\rho_{\rm d}$, and~$\rho_{\rm tot}$ are the gas, dust, and total density, respectively. That is, the gas-dust mixture is not locally isothermal, and its cooling time scale is finite. As noted above, the instability is suppressed by vertical buoyancy if the cooling time scale is too long. 
 
\subsection{Streaming instability} 
The streaming instability \citep{Youdin2005, Youdin2007a, Johansen2007a} results from the inwards radial drift of dust, which is caused by the difference in orbital velocity between the gas and the dust as well as their mutual coupling via drag. In contrast to the dust, the gas orbits with a sub-Keplerian velocity because it is supported against the radial stellar gravity by a pressure gradient. This pressure gradient constitutes a source of free energy that is tapped by the streaming instability \citep{Youdin2007a}. The linear growth rate of the instability is highest if the dust drift is fastest, that is when the dust stopping time is comparable to the inverse of the orbital frequency \citep{Weidenschilling1977}, and if the dust-to-gas density ratio is slightly greater than one \citep{Youdin2005}. Physical interpretations of the instability have been devised by \citet{Lin2017} and \citet{Squire2018}.

The turbulence that the streaming instability gives rise to in its non-linear regime can result in dust concentration in axisymmetric filaments. In these filaments, the dust accumulates in clumps that are sufficiently dense to collapse under their self-gravity and form planetesimals \citep{Johansen2007b, Bai2010b, Yang2014, Simon2016, Schaefer2017}. However, whether the streaming instability causes dust accumulation which is strong enough for planetesimal formation depends on the dust-to-gas surface density ratio and the size of the dust \citep{Johansen2009, Bai2010b, Carrera2015, Yang2017, Yang2018}, as well as the strength of the radial pressure gradient \citep{Bai2010c}. The required surface density ratio is higher than~$1\%$ -- the canonical value in the interstellar medium -- for all dust sizes. It can be enhanced sufficiently to reach the critical value globally by photoevaporation \citep{Carrera2017, Ercolano2017}, and locally in radial dust pile-ups \citep{Drazkowska2016} or at ice lines \citep{Ida2016, Schoonenberg2017, Schoonenberg2018, Drazkowska2017}.

Observations of binaries among the Trans-Neptunian objects provide evidence for planetesimal formation owing to the streaming instability: Their observed orbital inclinations relative to their heliocentric orbit as well as the ones that are found in simulations of binary formation by gravitational collapse are predominantly prograde \citep{Grundy2019, Nesvorny2010, Nesvorny2019}. In contrast, dynamical binary capture leads to either mostly retrograde inclinations or a similar number of prograde and retrograde ones, depending on the capture mechanism \citep{Schlichting2008}.

While it is considered to be and studied as one of the most promising mechanisms to induce planetesimal formation, the streaming instability has received comparably little attention as a process to drive turbulence, although this is its fundamental effect. 

Moreover, the streaming instability has been studied numerically almost exclusively in local shearing box simulations. Only \citet{Kowalik2013} present global two- and three-dimensional simulations, which reproduce the dust accumulation in dense axisymmetric filaments. 

In this paper, we study the streaming instability as a source of turbulence, employing axisymmetric global simulations with considerably larger domains than the ones that were simulated by \citet{Kowalik2013}. In contrast to these authors, we model the dust as Lagrangian particles rather than as a pressureless fluid, and take into account the vertical stellar gravity, which leads to the dust sedimenting to the mid-plane. We further apply adaptive mesh refinement to enhance the resolution of the dust mid-plane layer.

The paper is structured as follows: In Sect.~\ref{sect:model}, the simulations, their initial conditions, and the parameters that govern their evolution are described. In Sects.~\ref{sect:VSI},~\ref{sect:coexistence}, and~\ref{sect:SI}, respectively, we present our study of the turbulence in models of the vertical shear instability only, both the vertical shear and the streaming instability, and the streaming instability only. Implications and limitations of the study are discussed in Sect.~\ref{sect:discussion}. We summarize our results in Sect.~\ref{sect:summary}.

\section{Numerical model}
\label{sect:model}
We perform two-dimensional numerical simulations of the gas and the dust components of protoplanetary disks, including the mutual drag between the two components as well as the radial and the vertical stellar gravity. Magnetic fields, the self-gravity of the gas and the dust, and (non-numerical) viscosity are neglected. We employ version 4.5 of the adaptive-mesh refinement (AMR) finite volume code FLASH Code\footnote{\url{http://flash.uchicago.edu/site/flashcode/}} \citep{Fryxell2000}.

\subsection{Domains, boundary conditions, and resolutions}
\label{sect:domains}
The cylindrical, axisymmetric simulation domains extend from either~$1~\au$ to~$10~\au$ or from~$10~\au$ to~$100~\au$ in the radial dimension and~$1$ or~$2$ gas scale heights above and below the mid-plane in the vertical dimension. Since in our model the gas scale height increases non-linearly with the radius (see Eq.~\ref{eq:gas_scale_height}), the domains are shaped like isosceles trapezoids with curved rather than straight legs. To model this shape as accurately as possible with respect to the initial resolution, we initially create rectangular domains with a vertical size of two or four gas scale heights at the outer radial boundary. From these domains, we then exclude all blocks of~$10\times10$ grid cells whose distance to the mid-plane is greater than one or two local gas scale heights. For this purpose, we apply the obstacle block implementation that is included in the FLASH Code.\footnote{\label{footnote} Please address code requests regarding the modifications to the FLASH Code that we have implemented to conduct the simulations presented in this paper to \href{mailto:urs.schaefer@hs.uni-hamburg.de}{urs.schaefer@hs.uni-hamburg.de}. We note that we are not permitted to re-distribute the FLASH Code or any of its parts.}
%All modifications to the FLASH Code that we have implemented to conduct the simulations which are presented in this paper are available as a patch at ???. This patch may be used under the {\href{https://creativecommons.org/licenses/by-nc/4.0/}{Creative Commons Attribution-NonCommercial 4.0 International License}} if attribution to the authors and the title of this paper as well as the journal it is published in is maintained.}. 

Diode conditions are applied at both the radial and the vertical boundaries, i.e., the boundaries are permeable to gas and dust moving out of the domain, but reflect gas and dust that would move into it. At the vertical boundaries, the pressure in the guard cells is quadratically interpolated to maintain hydrostatic equilibrium. The temperature in the guard cells at all boundaries is reset to the initial value because we find this to be conducive to the stability of our simulations. In addition, the orbital velocity is corrected to account for the difference in temperature and stellar gravity between the guard and the non-guard cells\textsuperscript{\ref{footnote}}.

The FLASH Code employs the PARAMESH package \citep{MacNeice2000} to perform block-structured AMR. That is, the domain is subdivided into blocks of~$10\times10$ grid cells, which as a whole are refined or derefined if a refinement or derefinement criterion is fulfilled in any cell inside them. To resolve the gas and dust dynamics that are induced by the streaming instability in the disk mid-plane, we apply a (de-)refinement criterion which is based on the spatial dust distribution: The resolution is doubled if more than ten particles, which we use to model the dust, are located in one cell. On the other hand, the resolution is halved if no particles remain in a cell\textsuperscript{\ref{footnote}}. In addition, we initially increase the resolution by a factor of two or of four where the gas density exceeds~$1\%$ or~$10\%$, respectively, of the mid-plane density at the inner radial domain boundary.

In the domains with a radial size of~$9~\au$, the fiducial initial and maximum resolution amount to~$160$ and~$5\,120$ cells per~au, respectively, while in the domains with an extent of~$90~\au$ they are equal to~$10$ and~$320$ cells per~au, respectively. At the maximum resolution, this corresponds to more than~$200$ cells per gas scale height at all radii, which has been shown to be sufficient to resolve the formation of gravitationally unstable dust clumps owing to the streaming instability in local shearing box simulations \citep{Yang2014}. To investigate whether our findings are dependent on the resolution, we as well conduct simulations with a doubled initial and maximum resolution. 
 
\begin{table*}
\caption{Simulation parameters}
\centering
\resizebox{\hsize}{!}{
\begin{tabular}{lcccccccc}
\hline
\hline
Simulation name&Equation&$Z$ [\%]\tablefootmark{a}&Dust par-&$L_z$ [$H_{\rm g}$]\tablefootmark{c}&$L_r$ [$\au$]\tablefootmark{d}&$\Delta x_{\rm init}$ [$\au$]\tablefootmark{e}&$\Delta x_{\rm min}$ [$\au$]\tablefootmark{f}&$t_{\rm end}$ [$\kyr$]\tablefootmark{g}\\
&of state&&ticle size\tablefootmark{b}&&&&&\\
\hline
\hline
\textit{iso\_Lr=90au}\tablefootmark{h}&isothermal&-&-&$2$&$90$&$0.1$&$3.125\times10^{-3}$&$70$\\
\hline
\textit{iso\_Lr=90au\_Lz=4Hg}\tablefootmark{h}&isothermal&-&-&$4$&$90$&$0.1$&$3.125\times10^{-3}$&$70$\\
\hline
\textit{iso\_Lr=90au\_Lz=4Hg\_dou\_res}\tablefootmark{h}&isothermal&-&-&$4$&$90$&$0.05$&$1.56\times10^{-3}$&$70$\\
\hline
\hline
\textit{iso\_Z=0.02\_Lr=90au\_Lz=4Hg}&isothermal&$2$&$a=3~\cm$&$4$&$90$&$0.1$&$3.125\times10^{-3}$&$30$\\
\hline
\textit{iso\_Z=0.02\_Lr=90au\_Lz=4Hg\_tdinit=50kyr}\tablefootmark{i}&isothermal&$2$&$a=3~\cm$&$4$&$90$&$0.1$&$3.125\times10^{-3}$&$55$\\
\hline
\textit{iso\_Z=0.02\_Lr=90au\_Lz=4Hg\_tdinit=50kyr\_Hdinit=0.01Hginit}\tablefootmark{i,j}&isothermal&$2$&$a=3~\cm$&$4$&$90$&$0.1$&$3.125\times10^{-3}$&$55$\\
\hline
\textit{iso\_Z=0.04\_Lr=90au\_Lz=4Hg\_tdinit=50kyr}\tablefootmark{i}&isothermal&$4$&$a=3~\cm$&$4$&$90$&$0.1$&$3.125\times10^{-3}$&$55$\\
\hline
\textit{iso\_Z=0.1\_Lr=90au\_Lz=4Hg\_tdinit=50kyr}\tablefootmark{i}&isothermal&$10$&$a=3~\cm$&$4$&$90$&$0.1$&$3.125\times10^{-3}$&$55$\\
\hline
\hline
\textit{adi\_Lr=9au}\tablefootmark{h}&adiabatic&-&-&$2$&$9$&$6.25\times10^{-3}$&$1.95\times10^{-4}$&$0.3$\\
\hline
\textit{adi\_Z=0.02\_Lr=9au}&adiabatic&$2$&$a=3~\cm$&$2$&$9$&$6.25\times10^{-3}$&$1.95\times10^{-4}$&$0.3$\\
\textit{adi\_Z=0.02\_Lr=90au}&adiabatic&$2$&$a=3~\cm$&$2$&$90$&$0.1$&$3.125\times10^{-3}$&$2.5$\\
\hline
\textit{adi\_Z=0.02\_Lr=9au\_Hdinit=Hginit\tablefootmark{k}}&adiabatic&$2$&$a=3~\cm$&$2$&$9$&$6.25\times10^{-3}$&$1.95\times10^{-4}$&$0.4$\\
\textit{adi\_Z=0.02\_Lr=90au\_Hdinit=Hginit\tablefootmark{k}}&adiabatic&$2$&$a=3~\cm$&$2$&$90$&$0.1$&$3.125\times10^{-3}$&$2.5$\\
\hline
\textit{adi\_Z=0.01\_Lr=9au}&adiabatic&$1$&$a=3~\cm$&$2$&$9$&$6.25\times10^{-3}$&$1.95\times10^{-4}$&$0.3$\\
\textit{adi\_Z=0.01\_Lr=90au}&adiabatic&$1$&$a=3~\cm$&$2$&$90$&$0.1$&$3.125\times10^{-3}$&$2.5$\\
\hline
\textit{adi\_Z=0.02\_Lr=9au\_dou\_res}&adiabatic&$2$&$a=3~\cm$&$2$&$9$&$3.125\times10^{-3}$&$9.76\times10^{-5}$&$0.3$\\
\textit{adi\_Z=0.02\_Lr=90au\_dou\_res}&adiabatic&$2$&$a=3~\cm$&$2$&$90$&$0.05$&$1.56\times10^{-3}$&$2.5$\\
\hline
\textit{adi\_Z=0.02\_Lr=90au\_a=3mm}&adiabatic&$2$&$a=3~\mm$&$2$&$9$&$6.25\times10^{-3}$&$1.95\times10^{-4}$&$6$\\
\hline
\textit{adi\_Z=0.02\_Lr=90au\_taustop=0.1}&adiabatic&$2$&$\tau_{\rm stop}=0.1$&$2$&$90$&$0.05$&$1.56\times10^{-3}$&3.5\\
\hline
\hline
\end{tabular}
}
\tablefoot{
\tablefoottext{a}{Dust-to-gas surface density ratio.}
\tablefoottext{b}{Given either as a size~$a$ or as a dimensionless stopping time~$\tau_{\rm stop}$.}
\tablefoottext{c}{Vertical domain extent, where~$H_{\rm g}$ is the gas scale height. (Approximate value, see text.)}
\tablefoottext{d}{Radial domain size.}
\tablefoottext{e}{Initial grid cell edge length.}
\tablefoottext{f}{Minimum grid cell edge length (at maximum resolution).}
\tablefoottext{g}{Time after which simulation ends.}
\tablefoottext{h}{No dust particles included.}
\tablefoottext{i}{Dust particles initialized after~\mbox{$t_{\rm d,init}=50~\kyr$}.}
\tablefoottext{j}{Initial dust scale height equal to~$1\%$ of gas scale height.}
\tablefoottext{k}{Initial dust scale height equal to gas scale height.}

}
\label{table:simulations}
\end{table*}  
 
The simulation names and parameters are compiled in Table~\ref{table:simulations}. {Sets of simulations which are analyzed in different sections are separated by a double horizontal line. All names are composed of, in this the order and separated by underscores,
\begin{itemize}
\item the gas equation of state (isothermal or adiabatic),
\item the dust-to-gas surface density ratio (this is omitted if dust is not included in a simulation), and
\item the radial simulation domain size.
\end{itemize}
Where applicable, the names further indicate that
\begin{itemize}
\item the vertical domain extent amounts to four gas scale heights,
\item the resolution is twice the fiducial resolution,
\item the dust particles are introduced after~$50~\kyr$ rather than at the beginning of the simulation
\item the initial dust-to-gas scale height ratio is set to $1\%$ or~$100\%$, with the fiducial value being~$10\%$, and
\item the dust particle size deviates from the fiducial size of~$a=3~\cm$.
\end{itemize}
 
\subsection{Gas}
\label{sect:gas}
The equations of motion of the gas can be expressed as 
\begin{equation}
\frac{\del \rho_{\rm g}}{\del t}+\nabla\cdot\left(\rho_{\rm g} \bm{v}_{\rm g}\right)=0\text{ and}
\label{eq:continuity_equation}
\end{equation}
\begin{equation}
\frac{\del \bm{v}_{\rm g}}{\del t}+\left(\bm{v}_{\rm g}\cdot\nabla\right)\bm{v}_{\rm g}=-\frac{1}{\rho_{\rm g}}\nabla P-\nabla\Phi_{\rm S}+\frac{\rho_{\rm d}}{\rho_{\rm g}}\frac{\bm{v}_{\rm d}-\bm{v}_{\rm g}}{t_{\rm stop}},
\label{eq:momentum_equation}
\end{equation}
%\begin{equation}
%\msout{\frac{\del \rho_{\rm g} E_{\rm g}}{\del t}+\nabla\left[(\rho_{\rm g} E_{\rm g}+P)\bm{v}_{\rm g}\right]=-\rho_{\rm g}\bm{v}_{\rm g}\nabla\Phi_{\rm S}+\rho_{\rm d}\bm{v}_{\rm g}\frac{\bm{v}_{\rm d}-\bm{v}_{\rm g}}{t_{\rm stop}},}
%\label{eq:energy_equation}
%\end{equation}
where~$\bm{v}$ is the velocity. 
%,~\mbox{$E=\epsilon+1/2v^2$} the total specific energy, and~$\epsilon$ the internal specific energy.
The subscripts~${\rm g}$ and~${\rm d}$ refer to the gas and the dust, respectively. The stellar gravitational potential is given by~\mbox{$\Phi_{\rm S}=-GM_{\rm S}/\sqrt{r^2+z^2}$}, where~$G$ is the gravitational constant,~\mbox{$M_{\rm S}=1~M_{\odot}$} is the stellar mass,~and $r$ and~$z$, respectively, are the radial distance to the star and the height above or below the disk mid-plane. The last term on the right-hand side of Eq.~\ref{eq:momentum_equation} 
%and~\ref{eq:energy_equation}
results from the drag exerted by the dust on the gas, with~$t_{\rm stop}$ being the stopping time (see Sect.~\ref{sect:dust}).

%The system of equations is closed with the ideal gas equation of state,
%\begin{equation}
%\epsilon=\frac{P}{(\gamma-1)\rho_{\rm g}},
%\end{equation}
%where~$\gamma$ is the adiabatic index. We consider both an adiabatic and an isothermal equation of state. In the former case the vertical shear instability is stabilized by vertical buoyancy. In the latter case, that is to say for~\mbox{$\gamma\rightarrow1$}, the internal energy calculated from the above equation diverges. We therefore set~\mbox{$\gamma=5/3$} in both cases, but model a locally isothermal gas by continuously resetting the temperature to its initial value and computing the pressure from this temperature and the density. We explain in more detail how we implement both the isothermal and the adiabatic equation of state in Appendix~\ref{sect:equation_of_state}\textsuperscript{\ref{footnote}}.

To close the system of equations, we consider either an isothermal or an adiabatic equation of state. In the former case, the pressure is calculated as
\begin{equation}
P=\frac{RT}{\mu}\rho_{\rm g},
\end{equation}
where~$R$ is the ideal gas constant,~$T$ the temperature, and~\mbox{$\mu=2.33$} the mean molecular weight\textsuperscript{\ref{footnote}}. In this case, the adiabatic index~\mbox{$\gamma=1$}. To model the latter case, we employ a polytropic equation of state that is given by
\begin{equation}
P=K\rho_{\rm g}^{\gamma},
\end{equation}
where~\mbox{$K=RT\rho_{\rm g}^{1-\gamma}/\mu$} is the polytropic constant and the adiabatic index~\mbox{$\gamma=5/3$}\textsuperscript{\ref{footnote}}. While it is a local constant in time, the polytropic constant depends on the global temperature and density distributions (see Eqs.~\ref{eq:temperature},~\ref{eq:density}, and~\ref{eq:mid-plane_density}). If the gas is locally adiabatic, the vertical shear instability is stabilized by vertical buoyancy.

The initial temperature is adopted from the minimum mass solar nebula (MMSN) model \citep{Hayashi1981},
\begin{equation}
T=280~\K~\left(\frac{r}{1~\au}\right)^{-1/2}.
\label{eq:temperature}
\end{equation}
The steepness of this profile is in agreement with that of power-law fits to observed temperature distributions \citep{Andrews2005}. The radial temperature gradient gives rise a variation of the orbital speed with height. This vertical shear in turn is the source of energy of the vertical shear instability. 

The gas initially orbits with a sub-Keplerian velocity, which is determined by the balance between stellar gravity, centrifugal force, and pressure gradient. It is furthermore in vertical hydrostatic equilibrium. As it is vertically isothermal, the vertical density profile is thus given by
\begin{equation}
\rho_{\rm g}=\rho_{\rm g}(z=0)\exp\left[-\frac{\gamma GM_{\rm S}}{c_{\rm s}^2}\left(\frac{1}{r}-\frac{1}{\sqrt{r^2+z^2}}\right)\right],
\label{eq:density}
\end{equation}
where~\mbox{$c_{\rm s}=(\gamma RT/\mu)^{1/2}\propto r^{-1/4}$} (see Eq.~\ref{eq:temperature}) is the sound speed. 
%,~$R$ the ideal gas constant, and~\mbox{$\mu=2.33$} the mean molecular weight.
We set the initial mid-plane density to
\begin{equation}
\rho_{\rm g}(z=0)=10^{-9}~\g\,\cm^{-3}~\left(\frac{r}{1~\au}\right)^{-9/4}.
\label{eq:mid-plane_density}
\end{equation}

Numerically integrating over Eq.~\ref{eq:density} yields the surface density
\begin{equation}
\Sigma_{\rm g}=\int_{-1~H_{\rm g}}^{1~H_{\rm g}}\rho_{\rm g}~{\rm d}z=10^3~\g\,\cm^{-2}~\left(\frac{r}{1~\au}\right)^{-1},
\label{eq:surface_density}
\end{equation}
where
\begin{equation}
H_{\rm g}=\sqrt{\frac{c_{\rm s}^2r^3(2\gamma GM_{\rm S}-c_{\rm s}^2r)}{(c_{\rm s}^2r-\gamma GM_{\rm S})^2}}
\label{eq:gas_scale_height}
\end{equation}
is the gas scale height. This surface density profile is shallower than the one in the MMSN model \citep{Hayashi1981}, but consistent with that of observed young, massive protoplanetary disks \citep{Andrews2009, Andrews2010}.

We note that Eq.~\ref{eq:surface_density} gives the surface density as the density integrated over two gas scale heights, not integrated from~$-\infty$ to~$\infty$ as in the commonly used definition. Consequently, the total mass~$M_{\rm g,tot}$ in each of our domains depends not only on the radial, but also on the vertical domain extent:
\begin{equation}
M_{\rm g,tot}=
\begin{cases}
7.1\times10^{-3}~M_{\sun}~\left(\frac{L_r}{9~\au}\right)&L_z\approx2~H_{\rm g}\text{ or}\\
8.0\times10^{-3}~M_{\sun}~\left(\frac{L_r}{9~\au}\right)&L_z\approx4~H_{\rm g},
\end{cases}
\end{equation}
where~$L_r$ and~$L_z$ are the radial and vertical domain size, respectively.

The orbital speed can be expressed as~\mbox{$v_{{\rm g},\phi}=v_{\rm K}-\Pi c_{\rm s}$}, where
\begin{equation}
v_{\rm K}=\sqrt{\frac{GM_{\rm S}}{\left(r^2+z^2\right)^{3/2}}r^2}
\label{eq:Keplerian_speed}
\end{equation}
is the Keplerian speed. We adopt the dimensionless parameter~$\Pi$, which is introduced by \citep{Bai2010b}, to indicate the strength of the radial pressure gradient
\begin{equation}
\Pi=-\frac{1}{2c_{\rm s}\rho_{\rm g}\Omega_{\rm K}}\frac{{\rm d}P}{{\rm d}r},
\end{equation}
where~\mbox{$\Omega_{\rm K}=v_{\rm K}/r$} is the Keplerian orbital frequency. In the mid-plane, the parameter is equal to
\begin{equation}
\Pi(z=0)=4.6\times10^{-2}\frac{1}{\sqrt{\gamma}}\left(\frac{r}{1~\au}\right)^{1/4}
\label{eq:pressure_gradient}
\end{equation}
(see Eqs.~\ref{eq:temperature} and~\ref{eq:mid-plane_density}).

\subsection{Dust}
\label{sect:dust}
We model the dust as Lagrangian particles utilizing the massive active particle implementation that is included in the FLASH Code. We adopt an approach that is commonly used in local shearing box studies of the streaming instability \citep{Youdin2007a, Bai2010a}: The mass and momentum of every simulated particle is equal to the total mass and momentum of a large number of dust pebbles, while the drag coupling to the gas is the same as that of a single pebble. 

The dust is initially uniformly distributed in the radial dimension. The mass of the dust particles is determined by the their total number~\mbox{$N_{\rm d}=10^6$}, the dust-to-gas surface density ratio~$Z$, and the gas surface density. It can be expressed as
\begin{equation}
m_{\rm d}=\frac{1}{N_{\rm d}}\int_{L_r} 2\pi r\Sigma_{\rm d}~{\rm d}r=\frac{Z}{N_{\rm d}}\int_{L_r} 2\pi r\Sigma_{\rm g}~{\rm d}r.
\end{equation}
Because the gas surface density is inversely proportional to the radius (see Eq.~\ref{eq:surface_density}), the mass of all particles in a simulation is given by
\begin{equation}
\begin{split}
m_{\rm d}&=1.27\times10^{23}~\g~\left(\frac{Z}{0.01}\right)\left(\frac{L_r}{9~\au}\right)\\
&=0.14~M_{\rm Ceres}~\left(\frac{Z}{0.01}\right)\left(\frac{L_r}{9~\au}\right),
\end{split}
\end{equation}
where~\mbox{$M_{\rm Ceres}=9.3\times10^{23}~\g$} is the mass of Ceres. The simulated dust-to-gas surface density ratios range from~\mbox{$Z=0.01$} to~\mbox{$Z=0.1$}, with~\mbox{$Z=0.02$} being the fiducial value. We have verified that our results are converged with respect to the total number of particles by comparing simulations with~\mbox{$N_{\rm d}=5\times10^5$} and~\mbox{$N_{\rm d}=10^6$}.}

The initial vertical positions are randomly sampled from a Gaussian distribution with a scale height of~$10\%$ of the gas scale height. This scale height is in agreement with the thickness of the dust mid-plane layer observed by \citet{Pinte2016}. To investigate the dependence of our findings on the initial scale height, we additionally perform simulations with a dust-to-gas scale height ratio of~$0.01$ and of~$1$. The noise in the vertical distribution serves as a seed for the streaming instability. A comparison of simulations with two different vertical distributions has shown that our results do not noticeably depend on the random seed. 

We simulate dust with a fixed size of~\mbox{$a=3~\cm$} or of~\mbox{$a=3~\mm$}, or with a fixed dimensionless stopping time, which is equivalent to the Stokes number, of~\mbox{$\tau_{\rm stop}=0.1$}. Even if~\mbox{$a=3~\cm$}, the dust is smaller than the gas mean free path length at all densities in our model. Under this condition, that is to say in the Epstein regime, the dimensionless stopping time in the mid-plane is given by
\begin{equation}
\begin{split}
\tau_{\rm stop}(z=0)&=t_{\rm stop}(z=0)\Omega_{\rm K}(z=0)=\frac{a\rho_{\rm s}}{c_{\rm s}\rho_{\rm g}(z=0)}\Omega_{\rm K}(z=0)\\
&=6\times10^{-4}~\frac{1}{\sqrt{\gamma}}\left(\frac{a}{3~\mm}\right)\left(\frac{r}{1~\au}\right),
\end{split}
\label{eq:stopping_time}
\end{equation}
where~\mbox{$\rho_{\rm s}=1~\g\,\cm^{-3}$} is the dust material density (see also Eq.~\ref{eq:mid-plane_density}).

We note that the collisional growth to sizes greater than millimeters is prevented by dust grains bouncing or fragmenting under mutual collisions \citep{Guettler2010, Zsom2010, Birnstiel2011}, except for in the innermost regions of protoplanetary disks \citep{Birnstiel2012} and at ice lines \citep{Ros2013, Ros2019}. However, simulating centimeter-sized grains allows us to probe dimensionless stopping times of the order of~\mbox{$\tau_{\rm stop}=0.1$}. These are pertinent to a study of the turbulence that is driven by the streaming instability since the linear growth rate of the instability is highest if the dimensionless stopping time is close to one \citep{Youdin2005}.

The dust particles initially orbit with the Keplerian speed (see Eq.~\ref{eq:Keplerian_speed}). They are initialized either at the beginning of the simulations or after~$50~\kyr$. The latter is to give the vertical shear instability time to attain a saturated state before the introduction of the dust. To advance the particles in time, we adopt the Leapfrog algorithm for cylindrical geometries that has been devised by \citet{Boris1970}, as detailed in Appendix~\ref{sect:Leapfrog_algorithm}\textsuperscript{\ref{footnote}}. 

The implementation of the mutual drag between the gas and the dust is based on the cloud-in-cell mapping between the grid and the particles that is included in the FLASH Code. The algorithm can be described as follows: Firstly, the gas properties are mapped to the particles. Secondly, for each particle, the stopping time (see Eq.~\ref{eq:stopping_time}) and the change in velocity due to the drag exerted by the gas is computed. The corresponding change in particle momentum is then mapped to the grid. Finally, the change of the gas velocity in every grid cell is calculated from the particle momentum change. A more detailed description of the implementation can be found in Appendix~\ref{sect:drag}\textsuperscript{\ref{footnote}}.

\section{Vertical shear instability}
\label{sect:VSI}
In this section, we verify that our model can reproduce the findings of previous studies of the vertical shear instability, in particular the turbulent strength in its saturated state, despite the comparably small vertical domain sizes we consider. For this purpose, we analyze our model of locally isothermal gas in which dust is not included.

\begin{figure}[t]
\centering
\includegraphics[width=\columnwidth]{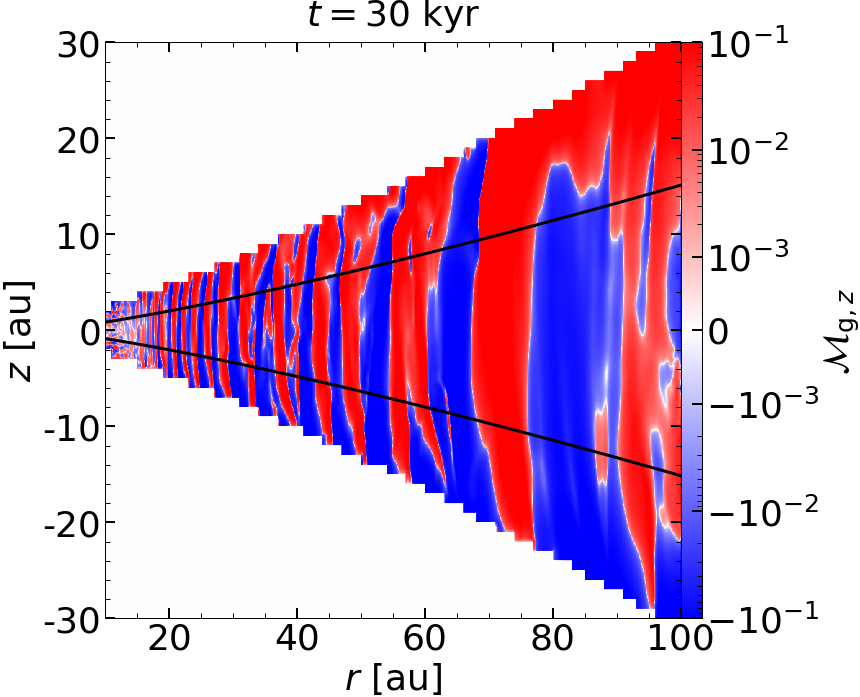}
\caption{Mach number of the vertical gas velocity~$\mathcal{M}_{{\rm g},z}$ as a function of radius~$r$ and height~$z$ in a simulation including a locally isothermal gas, but no particles (\textit{iso\_Lr=90au\_Lz=4Hg}). The simulation domain spans four gas scale heights in the vertical dimension. Black lines mark one gas scale height above and below the mid-plane. After~$30~\kyr$, the vertical shear instability has saturated at all radii and heights. It gives rise to perturbations which are characterized by a radial-to-vertical wavelength ratio much smaller than unity, by being bent outwards, and by a mirror symmetry relative to the mid-plane.}
\label{fig:Mach_number_no_dust_outer_region}
\end{figure}

In Fig.~\ref{fig:Mach_number_no_dust_outer_region}, the vertical gas motions in the model are depicted. The characteristic perturbations that the vertical shear instability gives rise to are reproduced. The perturbations are bent outwards, their radial wavelength is much less than their vertical wavelength, and they are symmetric with respect to the mid-plane in the saturated state of the instability (compare with, e.g., Figs.~2 and~3 of \citealt{Nelson2013} and Fig.~2 of \citealt{Stoll2014}).

%\begin{figure*}[t]
\begin{figure}[t]
%\centering
%\begin{minipage}{0.49\textwidth}
%\centering
%\includegraphics[width=\textwidth]{Mach_number_evolution_no_dust_20_au}
%\end{minipage}
%\hfill
%\begin{minipage}{0.49\textwidth}
%\centering
%\includegraphics[width=\textwidth]{Mach_number_evolution_no_dust_80_au}
%\end{minipage}
\centering
\includegraphics[width=\columnwidth]{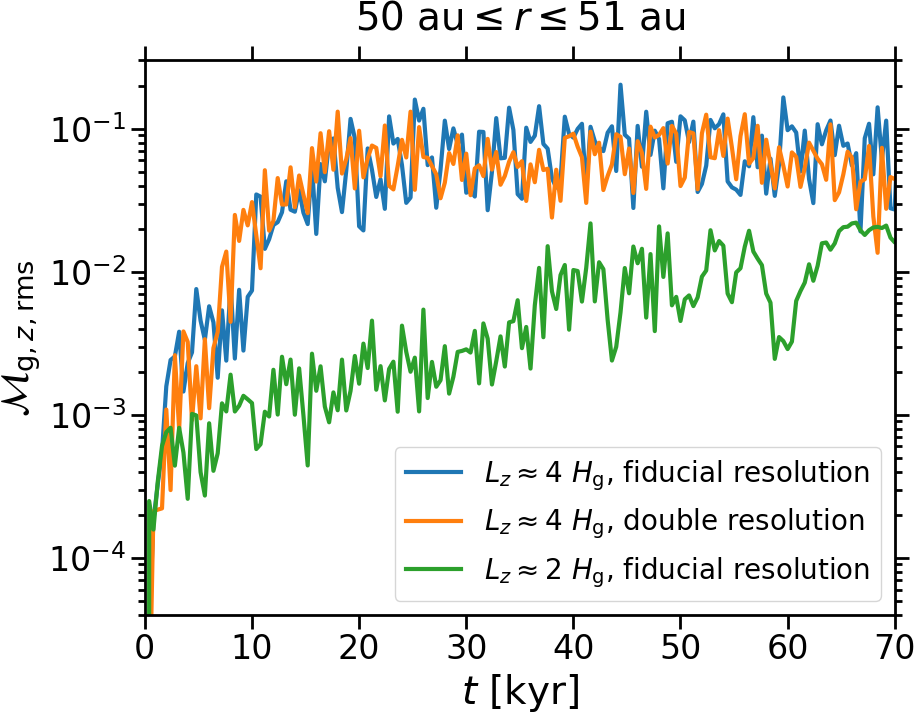}
\caption{
%Root mean square (RMS) of~$\mathcal{M}_{{\rm g},z}$ as a function og the time~$t$ in Keplerian orbital periods~$P_{\rm K}$. The root mean square is computed using the mass-weighted average over the vertical domain size and the radii given in the title of the respective panel. In the domain with a vertical extent of four gas scale heights (blue and orange lines), the Mach number saturates at~\mbox{$\mathcal{M}_{{\rm g},z,{\rm rms}}\approx10^{-2}-10^{-1}$}. This is consistent with the turbulent strength which \citet{Flock2017} find in their numerical study of the vertical shear instability. The Mach number remains lower until the simulations end in the vertically smaller domain with a size of two gas scale heights (simulation \textit{iso\_Lr=90au}; green line). Furthermore, it takes fewer orbits for the instability to reach a saturated state at~$80~\au\leq r\leq 81~\au$ (right panel) than at~$20~\au\leq r\leq 21~\au$ (left panel). At the latter radii, this state is attained earlier in the simulation with the doubled resolution (\textit{iso\_Lr=90au\_Lz=4Hg\_dou\_res}; orange line) than in the one with the fiducial resolution (\textit{iso\_Lr=90au\_Lz=4Hg}; blue line).
Root mean square (RMS) of~$\mathcal{M}_{{\rm g},z}$ as a function of the time~$t$. The root mean square is computed using the mass-weighted average over the vertical domain size and over~$1~\au$ extending from~\mbox{$r=50~\au$} to~$51~\au$. In the domain with a vertical dimension of four gas scale heights (blue and orange lines), the Mach number saturates at~\mbox{$\mathcal{M}_{{\rm g},z,{\rm rms}}\approx10^{-1}$}. This value is consistent with the one which \citet{Flock2017} find in their numerical study of the vertical shear instability. The turbulent strength is similar in the simulation with the fiducial resolution (\textit{iso\_Lr=90au\_Lz=4Hg}; blue line) and the one with the doubled resolution (\textit{iso\_Lr=90au\_Lz=4Hg\_dou\_res}; orange line). However, it remains lower until the simulations end in the domain with a vertical size of two gas scale heights (simulation \textit{iso\_Lr=90au}; green line).
}
\label{fig:Mach_number_evolution_no_dust}
%\end{figure*}
\end{figure}

We find that the strength of the turbulence which the vertical shear instability induces depends on the vertical domain extent. This can be seen from Fig.~\ref{fig:Mach_number_evolution_no_dust}, in which we depict the vertical gas velocity in our models with a vertical domain size of two and of four gas scale heights. Throughout the simulations, the turbulence is considerably weaker in the vertically smaller domain than in the vertically larger one. The reason for this is likely that the vertical shear at the vertical domain boundaries is less in smaller domains, which entails a decline of the linear growth rate of the surface modes of the instability \citep{Lin2015}.

In the vertically larger domain, the vertical shear instability saturates at a Mach number of the vertical gas velocity of~\mbox{$\mathcal{M}_{{\rm g},z}\approx0.1$}. This value does not depend significantly on the resolution. It is consistent with the Mach number which \citet{Flock2017} find. This is regardless of these authors simulating a domain with an aspect ratio of~\mbox{$z/r=0.35$}, while our domain extends to between~\mbox{$z/r=0.17$} at~\mbox{$r=10~\au$} and~\mbox{$z/r=0.3$} at~\mbox{$r=100~\au$}. Furthermore, they employ a radiative transfer model rather than assuming the gas to be locally isothermal as we do.

Therefore, in the following we investigate the vertical shear instability and its coexistence with the streaming instability using our model with a vertical domain size to two gas scale heights above and below the mid-plane. 

\section{Coexistence of vertical shear instability and streaming instability}
\label{sect:coexistence}
Using our model with dust and an isothermal gas equation of state, we investigate how the vertical shear instability and the streaming instability interact. We consider two scenarios: In the scenario \textit{SIwhileVSI}, the streaming instability grows simultaneously with the vertical shear instability. In the scenario \textit{SIafterVSI}, on the other hand, the streaming instability is not active before the vertical shear instability has saturated. We model the latter scenario by introducing the dust into the simulations after~$50~\kyr$. At this point, the vertical shear instability has reached a saturated state at all radii (see Fig.~\ref{fig:Mach_number_no_dust_outer_region}).

\subsection{Source of turbulence}
\begin{figure*}[t]
\centering
\begin{minipage}{0.49\textwidth}
\centering
\includegraphics[width=\textwidth]{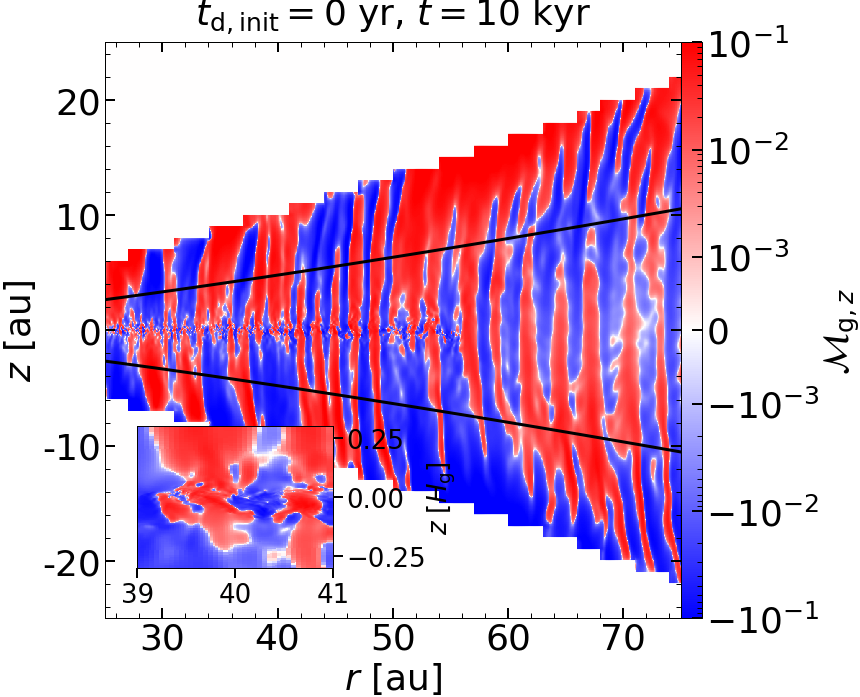}
\end{minipage}
\hfill
\begin{minipage}{0.49\textwidth}
\centering
\includegraphics[width=\textwidth]{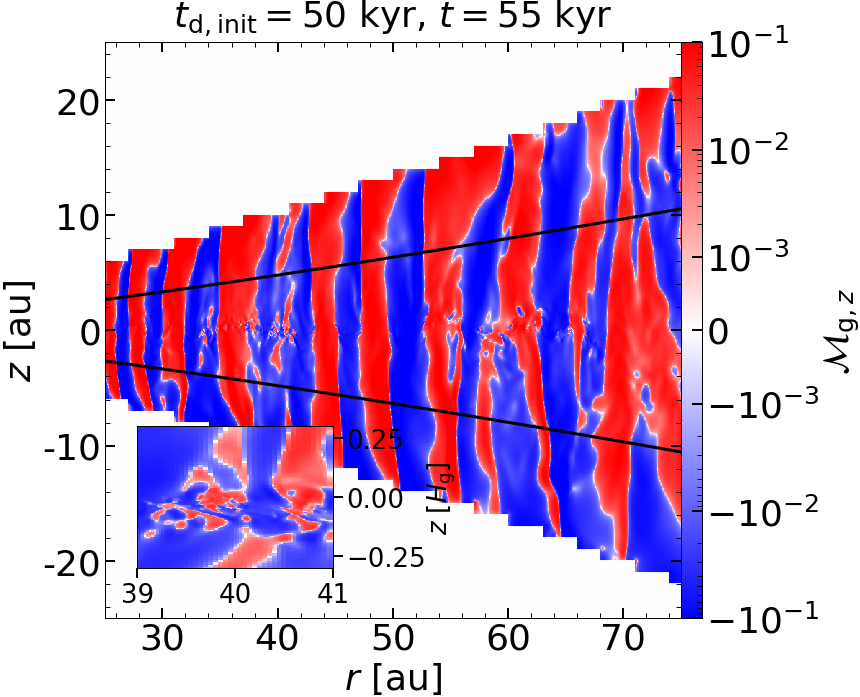}
\end{minipage}
\caption{Mach number~$\mathcal{M}_{{\rm g},z}$ as a function of~$r$ and~$z$ in two simulations with dust and an isothermal gas equation of state. As in their dust-free equivalent (compare with Fig.~\ref{fig:Mach_number_no_dust_outer_region}), the vertical shear instability induces large-scale perturbations away from the mid-plane. 
%(Black lines indicate where~$|z|$ is equal to one gas scale height.)
In the simulation \textit{iso\_Z=0.02\_Lr=90au\_Lz=4Hg} (left panel), in which we initialize the dust at the beginning, at all radii in the mid-plane small-scale perturbations can be seen. These perturbations are as well observable in the simulation \textit{iso\_Z=0.02\_Lr=90au\_Lz=4Hg\_tdinit=50kyr} (right panel), in which the dust is introduced after~$50~\kyr$, even though not at all radii and extending to greater heights. The inlays with a vertical extent of~$0.6$ gas scale heights show the enlarged perturbations. However, such small-scale perturbations do not exist in the corresponding simulation without dust.}
\label{fig:Mach_number_dust_outer_region}
\end{figure*}

Fig.~\ref{fig:Mach_number_dust_outer_region} shows the Mach number of the vertical gas motions in the scenario \textit{SIwhileVSI} (left panel) and the scenario \textit{SIafterVSI} (right panel). In both cases, away from the mid-plane the vertical shear instability is the primary source of turbulence and induces the characteristic large-scale perturbations (compare with Fig.~\ref{fig:Mach_number_no_dust_outer_region}). 

In the mid-plane, however, small-scale perturbations can be seen at all radii in the former scenario. Similar perturbations exist in the latter scenario, although only at certain radii and not as limited in vertical extent. These small-scale perturbations are not present in our model of the vertical shear instability only. They resemble the perturbations that \citet{Li2018} observe in the mid-plane of their local shearing box simulations of the streaming instability (see their Fig. 2).

\begin{figure*}[t]
\centering
\begin{minipage}{0.49\textwidth}
\centering
\includegraphics[width=\textwidth]{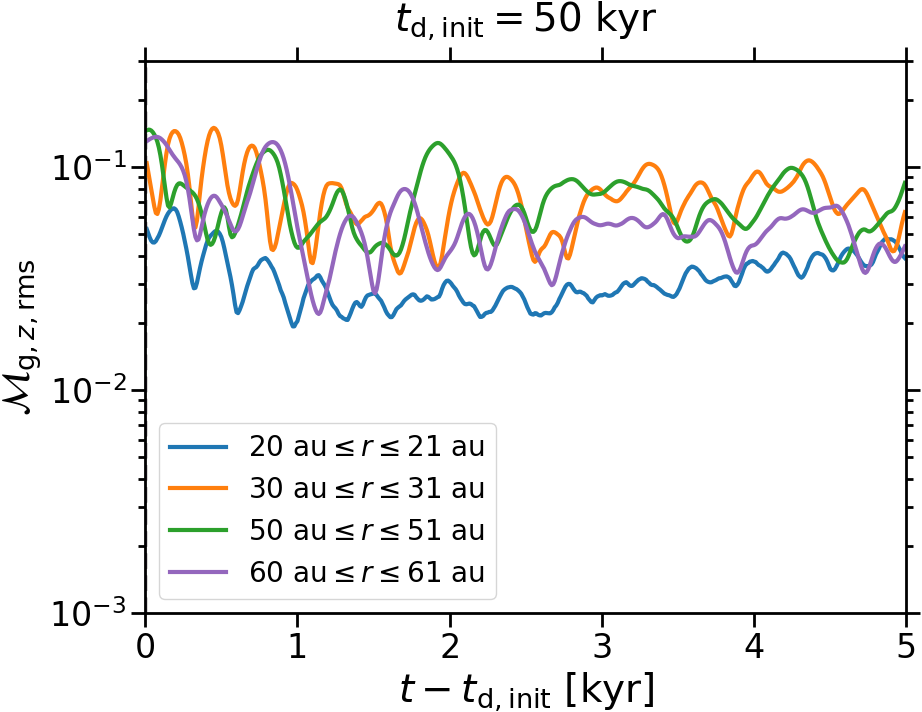}
\end{minipage}
\hfill
\begin{minipage}{0.49\textwidth}
\centering
\includegraphics[width=\textwidth]{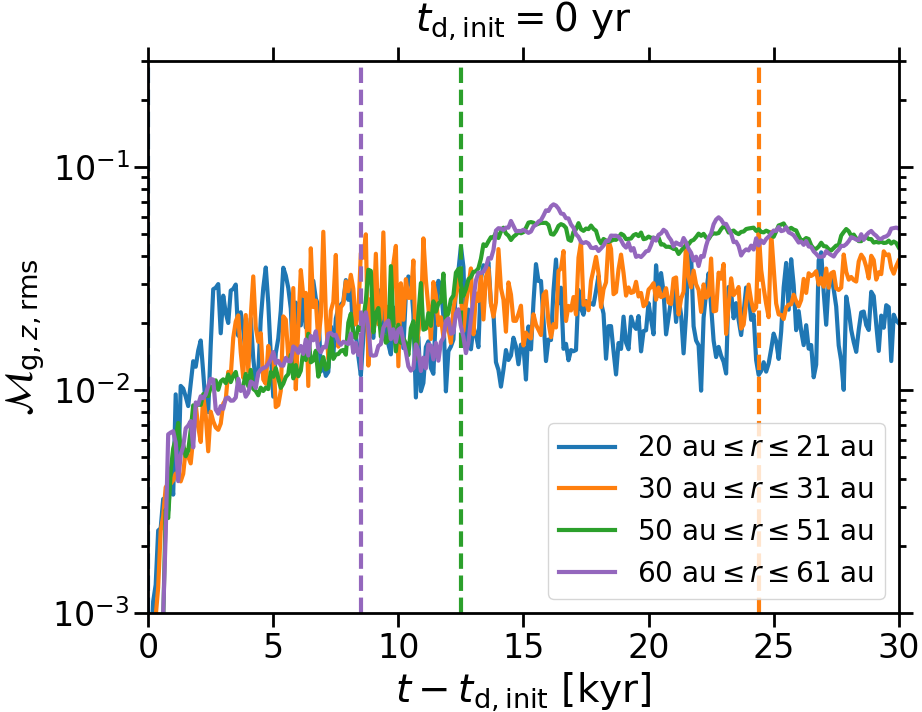}
\end{minipage}
\caption{RMS of~$\mathcal{M}_{{\rm g},z}$, averaged over the vertical domain extent and the radii given in the legend, as a function of~$t$ after the initialization of the dust at~$t_{\rm d,init}$. The mean is weighted by mass, i.e., more weight is assigned to the Mach number in the mid-plane than to the one away from it. In the simulation \textit{iso\_Z=0.02\_Lr=90au\_Lz=4Hg\_tdinit=50kyr} (left panel), the vertical shear instability has saturated before the dust is introduced. The Mach number remains about constant at a value~\mbox{$\mathcal{M}_{{\rm g},z,{\rm rms}}\approx10^{-1}$} after the dust initialization (compare with Fig.~\ref{fig:Mach_number_evolution_no_dust}). That is, in this simulation the vertical shear instability is the primary source of turbulence in the mid-plane. In contrast, in the simulation \textit{iso\_Z=0.02\_Lr=90au\_Lz=4Hg} (right panel), in which the vertical shear instability and the streaming instability begin to grow at the same time, the Mach number saturates at a lower value of~\mbox{$\mathcal{M}_{{\rm g},z,{\rm rms}}\approx10^{-2}$}. After no dust remains at a radius because of its radial drift (this point is marked with a dashed line), the Mach number increases until it is approximately equal to the value at the same radius in the simulation \textit{iso\_Z=0.02\_Lr=90au\_Lz=4Hg\_tdinit=50kyr}. We find that, as long as dust is present, the streaming instability drives the turbulence in the mid-plane in the simulation \textit{iso\_Z=0.02\_Lr=90au\_Lz=4Hg}.}
\label{fig:Mach_number_evolution_dust}
\end{figure*}

In Fig.~\ref{fig:Mach_number_evolution_dust}, we compare the time-dependence of the Mach number in the two scenarios. The mass-weighted vertical average of the Mach number is depicted, i.e., the turbulent strength in the mid-plane is weighted more heavily than the strength away from it. In the scenario \textit{SIafterVSI}, we find that the initialization of the dust does not cause the Mach number to deviate significantly from the value of~\mbox{$\mathcal{M}_{{\rm g},z}\approx0.1$} that the vertical shear instability induces after it has saturated (compare with Fig.~\ref{fig:Mach_number_evolution_no_dust}).

In contrast, in the scenario \textit{SIwhileVSI} the Mach number saturates at a lower value of~\mbox{$\mathcal{M}_{{\rm g},z}\approx0.01$.} In the local shearing box simulations without vertical stellar gravity which are presented by~\citet{Johansen2007a}, the streaming instability drives turbulence with a similar strength. After saturation, the Mach number does not vary significantly until, owing to its radial drift, locally no dust is left. Subsequently, it increases to approximately the value the vertical shear instability in its saturated state gives rise to in the scenario \textit{SIafterVSI}.

\begin{figure}[t]
\centering
\includegraphics[width=\columnwidth]{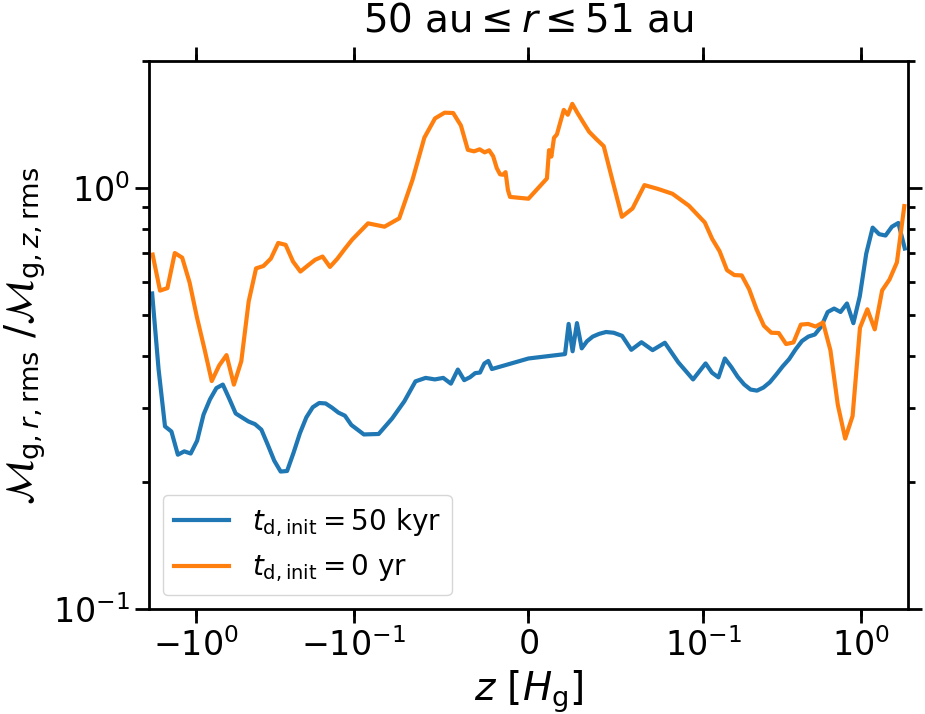}
\caption{Ratio of the RMS Mach number of the radial gas velocity $\mathcal{M}_{{\rm g},r}$ to the RMS of~$\mathcal{M}_{{\rm g},z}$ as a function of~$z$ in units of gas scale heights. The abscissa is scaled linearly between~$-0.1$ and~$0.1$ gas scale heights and logarithmically otherwise. To compute the RMS values, we take the mass-weighted average over~\mbox{$r=50~\au$} to~$51~\au$ and the average over~$1~\kyr$ through~\mbox{$t=55~\kyr$} in the case of the simulation \textit{iso\_Z=0.02\_Lr=90au\_Lz=4Hg\_tdinit=50kyr} (blue line) and through~\mbox{$t=10~\kyr$} in the case of the simulation \textit{iso\_Z=0.02\_Lr=90au\_Lz=4Hg} (orange line). Away from the mid-plane, the radial-to-vertical Mach number ratio is less than unity in both simulations. While this anisotropy extends over all heights if the dust is introduced after~$50~\kyr$, the Mach number ratio is close to one in the mid-plane if the dust is initialized at the start. This is indicative of the vertical shear instability being the primary source of turbulence in the mid-plane in the former case, but the streaming instability in the latter case.}
\label{fig:Mach_number_ratio}
\end{figure}

Fig.~\ref{fig:Mach_number_ratio} shows the ratio of the radial to the vertical Mach number in both scenarios. The turbulent velocity in the radial direction is significantly less than the one in the vertical direction in the scenario \textit{SIafterVSI}. This is consistent with \citet{Stoll2016} and \citet{Stoll2017} showing that the vertical shear instability drives anisotropic turbulence. On the other hand, in the scenario \textit{SIwhileVSI} the radial and the vertical Mach number are comparable near and in the mid-plane. \citet{Johansen2007a} find the streaming instability to cause isotropic turbulence.

From the fact that the turbulence is comparably weak and isotropic as well as from the presence of small-scale perturbations that are not observable in our model of the vertical shear instability only, we conclude that the streaming instability is the primary source of turbulence in the dust mid-plane layer if it starts to operate at the same time as the vertical shear instability. One possible explanation for this is that the streaming instability grows faster in turbulent strength than the vertical shear instability. This is evident when comparing the time which the vertical shear instability (see Fig.~\ref{fig:Mach_number_evolution_no_dust}) and the streaming instability (see the left panel of Fig.~\ref{fig:Mach_number_evolution_dust}) take to saturate.

The turbulence is mainly driven by the vertical shear instability, however, if it has attained a saturated state before the streaming instability can begin to grow. In this state, we find it to give rise to turbulence with a higher Mach number than the streaming instability in the scenario \textit{SIwhileVSI}. We note, however, that the small-scale perturbations, which we find to be induced by the streaming instability at all radii in the scenario \textit{SIwhileVSI}, are observable locally also in the scenario \textit{SIafterVSI} (see the right panel Fig.~\ref{fig:Mach_number_dust_outer_region}).

The vertical shear instability would likely be the main source of turbulence in the scenario \textit{SIwhileVSI} as well if the dust grains were too small to settle and trigger the streaming instability before the vertical shear instability has saturated. Equating a saturation time scale of~$30$ orbital periods \citep{Stoll2014} with the settling time scale as given by Eq.~10 of \citet{Chiang2010} yields a critical dimensionless dust stopping time of~${\sim}0.005$. However, since simulating such small dust grains as particles is computationally very expensive, we do not explore this scenario.

\subsection{Dependence of turbulent strength on dust density}
As explained in Sect.~\ref{sect:introduction}, settling dust introduces an effective vertical gas buoyancy that can suppress the vertical shear instability. This is a second possible reason -- besides the more rapid growth in turbulent strength of the streaming instability -- for the streaming instability being the source of turbulence in the dust layer in the scenario \textit{SIwhileVSI}.

\citet{Lin2019} performs simulations of the vertical shear instability which include dust from the beginning, as in the scenario \textit{SIwhileVSI}. They show that the dust-induced buoyancy leads to the Mach number in the mid-plane decreasing from~\mbox{$\mathcal{M}_{{\rm g},z}\approx10^{-1}$} to~\mbox{$\mathcal{M}_{{\rm g},z}\approx10^{-2}$} if the dust-to-gas surface density or the stopping time of the dust exceed a threshold value. These threshold values are correlated: For a surface density ratio of~$1\%$, the critical dimensionless stopping time of the dust is equal to~$0.005$ (see their Fig.~5). For a stopping time of~$0.001$, the critical surface density ratio amounts to~$3\%$ (see their Fig.~9). This is consistent with the Mach number of~\mbox{$\mathcal{M}_{{\rm g},z}\approx10^{-2}$} we find in the scenario \textit{SIwhileVSI}, in which the surface density ratio is equal to~$2\%$\footnote{The loss of gas mass through the domain boundaries of our simulations leads to an increase of the dust-to-gas surface density ratio with time. This effect is negligible if we simulate the dust over a time span of~$5~\kyr$ or less (see Table~\ref{table:simulations}). It is significant, however, in the simulation \textit{iso\_Z=0.02\_Lr=90au\_Lz=4Hg}, which ends after~$30~\kyr$. At this point, the surface density ratio has increased to~$2.6\%$.} and the stopping time of the dust ranges from~$0.046$ to~$0.46$.

However, while \citet{Lin2019} associates the turbulence in the mid-plane of their simulations with the vertical shear instability, we find it to be driven by the streaming instability if both instabilities begin to operate at the same time. \citet{Lin2019} notes that they apply a diffusive numerical scheme which probably suppresses the growth of the streaming instability. In addition, they model tightly coupled dust and gas employing a one-fluid approach.

\begin{figure}[t]
\centering
\includegraphics[width=\columnwidth]{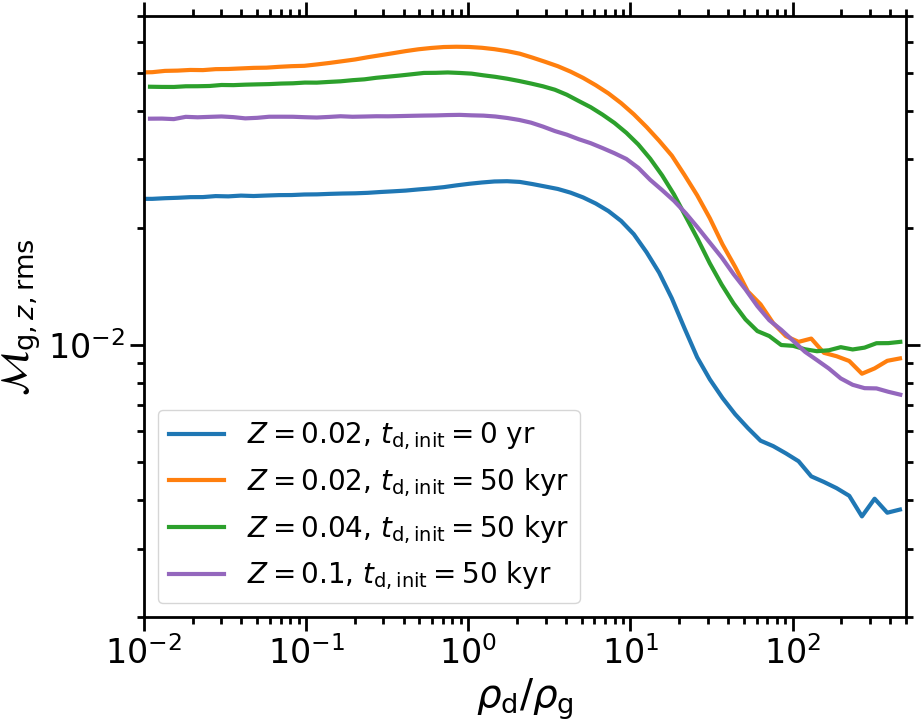}
\caption{RMS of~$\mathcal{M}_{{\rm g},z}$ as a function of the dust-to-gas density ratio~$\rho_{\rm d}/\rho_{\rm g}$. The orange, green, and purple lines represent the turbulent strength in simulations in which we initialize the dust after the vertical shear instability has saturated. The strength at low to intermediate volume density ratios in the three simulations is marginally smaller if the dust-to-gas surface density ratio is greater. 
%(The surface density ratios~$Z$ are given the legend.)
Nevertheless, the Mach number is higher in all of these simulations than in the simulation in which we introduce the dust at the start (blue line). We find that the stronger turbulence in the former simulations is driven by the vertical shear instability, while the weaker turbulence in the latter one is caused by the streaming instability. Independent of the source of the turbulence, its strength decreases at dust densities larger than the gas density in all four simulations.}
\label{fig:Mach_number_density}
\end{figure}

To investigate whether the vertical shear instability is quenched by the dust-induced buoyancy in the scenario \textit{SIafterVSI}, we analyze how the turbulent strength depends on the dust density. We note that this analysis does not involve information about the spatial distribution of the dust. That is, the strength at a certain dust density is the sum of the bulk motions of regions with this density and the internal turbulence in these regions. 

In Fig.~\ref{fig:Mach_number_density}, the Mach number is shown as a function of the dust-to-gas density ratio in both scenarios. Overall, the Mach number is higher in the scenario \textit{SIafterVSI} (orange, green, and purple lines) than in the scenario \textit{SIwhileVSI} (blue line). This is consistent with stronger turbulence in the dust layer being induced by the vertical shear instability in the former scenario, and weaker turbulence by the streaming instability in the latter scenario. 

In the scenario \textit{SIafterVSI}, at low to intermediate dust-to-gas volume density ratios the Mach number decreases with increasing dust-to-gas surface density ratio, although only slightly. This can be explained by the vertical shear instability being suppressed if the surface density ratio, i.e., the total dust mass, is greater. Then, the vertical bulk motions of regions with these volume density ratios are weaker.

Furthermore, the vertical shear instability is gradually quenched if the dust density increases to values greater than the gas density. The Mach number is as low as~\mbox{$\mathcal{M}_{{\rm g},z}\approx0.01$} for the highest density ratios. This turbulent strength is similar to the one which the streaming instability causes globally in the dust layer in the scenario \textit{SIwhileVSI}. We note, though, that even in the regions with the highest density ratios in the scenario \textit{SIafterVSI}, the Mach number is greater than in the regions with the same density ratios in the scenario \textit{SIwhileVSI}. This is because the vertical shear instability gives rise to stronger bulk motions of these regions in the former scenario than the streaming instability in the latter scenario.

The turbulent strength that the streaming instability induces in the scenario \textit{SIwhileVSI} is as well lower if the density ratio exceeds a few. This is in agreement with \citet{Johansen2009} finding that the collision speeds of dust grains in the filaments forming in their simulations of the streaming instability are smaller if the dust density is higher.

\subsection{Vertical and radial dust concentration}
\begin{figure*}[t]
\centering
\begin{minipage}{0.49\textwidth}
\centering
\includegraphics[width=\textwidth]{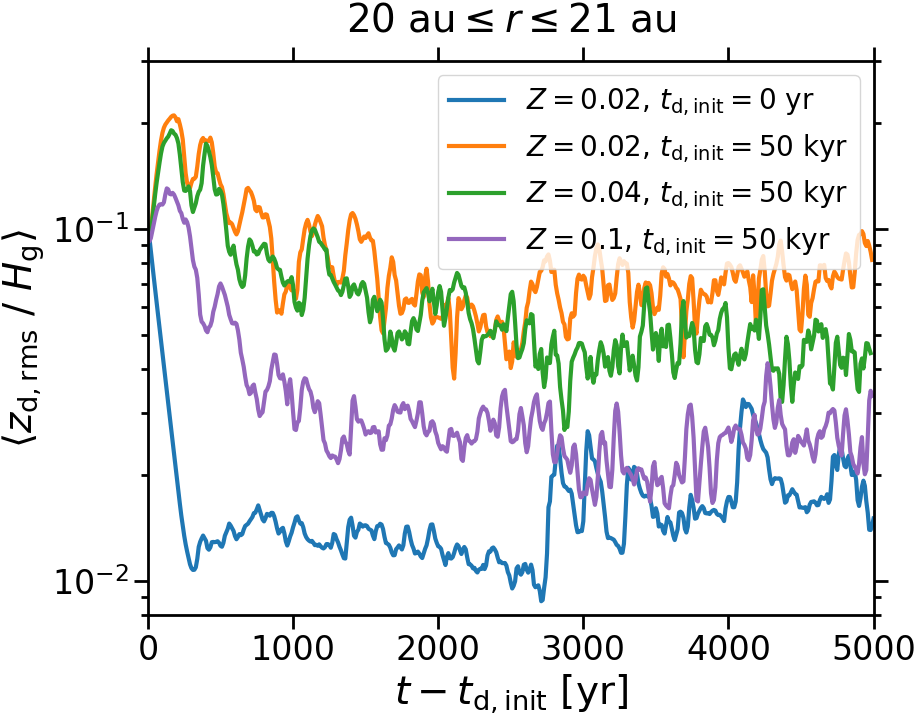}
\end{minipage}
\hfill
\begin{minipage}{0.49\textwidth}
\centering
\includegraphics[width=\textwidth]{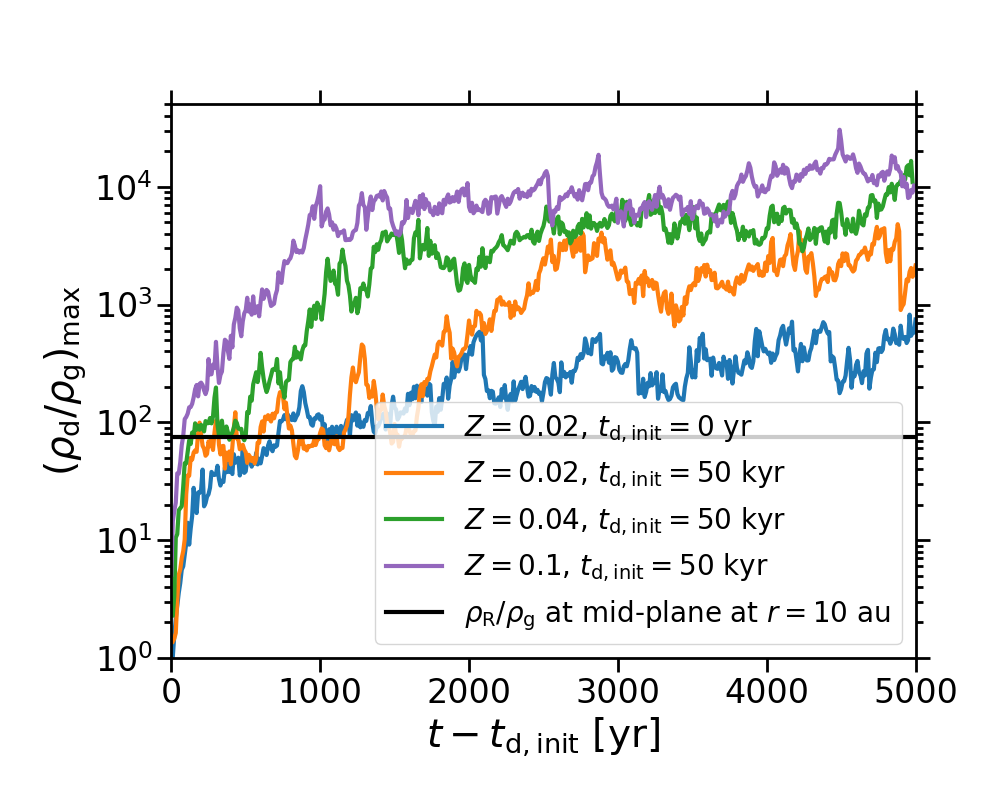}
\end{minipage}
\caption{Ratio of the dust scale height to the gas scale height~$H_{\rm g}$ (left panel) and maximum of the dust-to-gas density ratio~$\rho_{\rm d}/\rho_{\rm g}$ (right panel) as functions of~$t$ after the dust is introduced at~$t_{\rm d,init}$. We compute the dust scale height as the RMS of the vertical particle positions~$z_{\rm d}$, and average the scale height ratio over~$1~\au$ spanning from~\mbox{$r=20~\au$} to~$21~\au$. For a dust-to-gas surface density ratio of~\mbox{$Z=2\%$}, the dust scale height amounts to~${\sim}1\%$ of the gas scale height if the streaming instability induces the vertical diffusion of the dust (blue line). On the other hand, it is equal to~${\sim}10\%$ for the same surface density ratio if the diffusion is caused by the vertical shear instability (orange line). The scale height which is induced by the vertical shear instability decreases with increasing surface density ratio (green and purple line). It is higher than the value that the streaming instability gives rise to, though, for all surface density ratios that we consider. In contrast, the maximum dust volume density is significantly smaller if the streaming instability drives the turbulence in the dust layer than if the turbulence is caused by the vertical shear instability. In the former case, the dust-to-gas volume density ratio amounts to a few~$100$ for a surface density ratio of~$2\%$. In the latter case, on the other hand, it exceeds~$10^3$ for the same surface density ratio, and~$10^4$ for a surface density ratio of~$10\%$. In all cases, the maximum dust density is greater than the Roche density~$\rho_R$ in the mid-plane at the inner radial boundary of the simulation domains, i.e., the maximum Roche density in the simulations. The ratio of this Roche density to the gas density is marked as a black line.}
\label{fig:scale_height_density_maximum}
\end{figure*}

The dust-to-gas scale height ratio in the scenarios \textit{SIwhileVSI} (blue line) and \textit{SIafterVSI} (orange, green, and purple lines) is shown the left panel of Fig.~\ref{fig:scale_height_density_maximum}. In the former scenario, the streaming instability is the source of turbulence in the dust layer. The dust sedimentation and the vertical diffusion of the dust which it causes reach a balance at a dust-to-gas scale height ratio of~${\sim}1\%$ (see also Sect.~\ref{sect:SI}). 

Since the vertical shear instability drives stronger turbulence than the streaming instability, we find the equilibrium dust scale height to be greater in the scenario \textit{SIafterVSI}. On the other hand, it decreases with increasing dust-to-gas surface density ratio in this scenario: For a ratio of~$2\%$, the dust scale height amounts to~${\sim}10\%$ of the gas scale height. In comparison, if the ratio is equal to~$10\%$ the scale height is close to the value in the scenario \textit{SIwhileVSI}. We show in Sect.~\ref{sect:SI} that the scale height which is induced by the streaming instability is largely independent of the surface density ratio.

The dust settling to smaller scale heights for higher surface density ratios is most probably a consequence of the vertical shear instability being more suppressed by the dust-induced buoyancy. \citet{Lin2019} finds the instability to diffuse dust with a dimensionless stopping time of~$0.001$ to a scale height that is similar to the gas scale height if the surface density ratio amounts to~$1\%$, but is an order of magnitude smaller if the ratio is equal to~$5\%$ (see their Fig. 9). These scale heights are significantly greater than the ones in the scenario \textit{SIafterVSI}. This is likely because we simulate dust with stopping times between~$0.046$ to~$0.46$, which is more weakly coupled to the gas and is therefore less elevated by the vertical gas motions.

To investigate whether in the scenario \textit{SIafterVSI} the vertical shear instability can be quenched if the dust is initialized with a smaller scale height and thus a higher mid-plane density, we conduct a simulation with an initial surface density ratio of~$2\%$ and an initial dust scale height of~$1\%$ of the gas scale height rather than the fiducial value of~$10\%$. Despite this scale height being comparable to the one which is induced by the streaming instability in the scenario \textit{SIwhileVSI} and the initial mid-plane dust-to-gas density ratio being of order unity, we find that the vertical shear instability is not noticeably affected. The dust is elevated to a dust-to-scale height ratio of~${\sim}10\%$ in less than an orbital period in this simulation.

In the right panel of Fig.~\ref{fig:scale_height_density_maximum}, we depict the maximum dust-to-gas density ratio in the two scenarios. We note that this maximum is stochastic and dependent on the resolution \citep{Johansen2007a, Bai2010c}. Nonetheless, we find the maximum dust density to exceed the maximum Roche density
\begin{equation}
\rho_{\rm R,max}=\frac{9\Omega_{\rm K}(r=10~\au,~z=0)}{4\pi G}
\end{equation}
in both scenarios. That is, if the dust self-gravity were included in our model, local dust overdensities could undergo gravitational collapse and form planetesimals. This is in line with expectations for the scenario \textit{SIwhileVSI} since the surface density ratio of~$2\%$ in this scenario exceeds the critical value for the dust concentration by the streaming instability to be strong enough to lead to planetesimal formation \citep{Carrera2015, Yang2017}.

If the radial concentration of the dust were comparable in both scenarios, the maximum dust density would by tendency be greater in the scenario \textit{SIwhileVSI}, as the dust scale height is smaller in this scenario. On the contrary, we find the maximum dust density to be considerably higher in the scenario \textit{SIafterVSI}. The maximum of the density ratio amounts to a few hundred in the scenario \textit{SIwhileVSI}, but to a few thousand for the same surface density ratio in the scenario \textit{SIafterVSI}, and to more than~$10^4$ for higher surface density ratios. We investigate the radial dust concentration in the following.

\begin{figure*}[t]
\centering
\includegraphics[width=\textwidth]{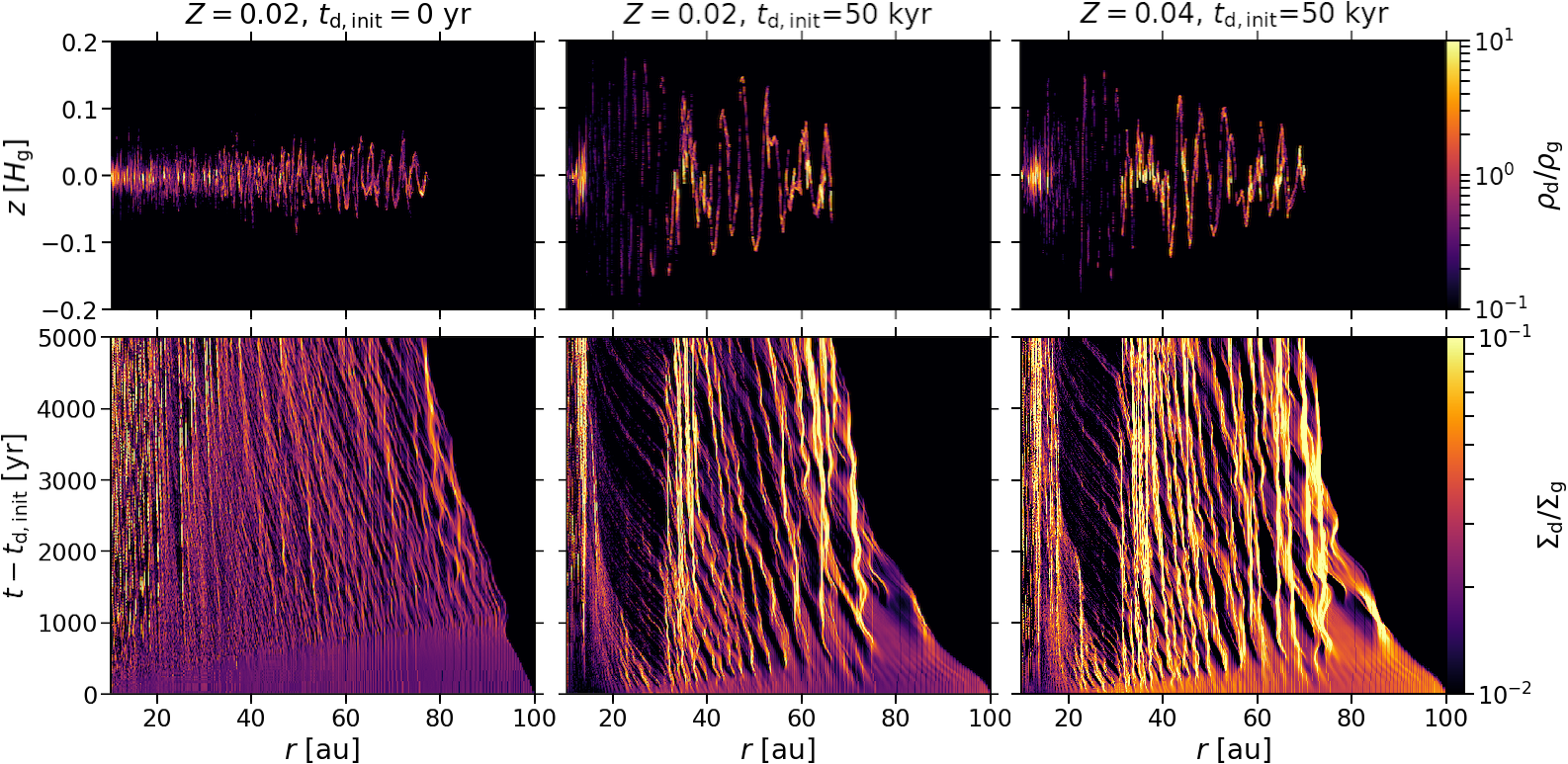}
\caption{Dust-to-gas volume density ratio~$\rho_{\rm d}/\rho_{\rm g}$ as a function of~$r$ and~$z$ (upper panels) as well as dust-to-gas surface density ratio~$\Sigma_{\rm d}/\Sigma_{\rm g}$ as functions of~$r$ and~$t$ (lower panels). The upper panels show the spatial dust distribution~$5~\kyr$ after the dust initialization. We compare a simulation in which the turbulence in the dust layer is driven by the streaming instability (left panels) and two simulations in which the vertical shear instability is the main source of turbulence (middle and right panels). For the same dust-to-gas surface density ratio of~$Z=0.02$, the dust scale height is smaller, but the radial dust concentration is weaker in the former case. (The values of~$Z$ are specified in the titles.) In the latter case, more dust is accumulated in overdensities if the surface density ratio is higher. Some of the accumulations are sufficiently dense for their radial drift to cease almost entirely.}
\label{fig:density}
\end{figure*}

In the upper panels of Fig.~\ref{fig:density}, we show the spatial dust distribution~$5~\kyr$ after the dust is introduced in the scenario \textit{SIwhileVSI} (left panel) and in the scenario \textit{SIafterVSI} (middle and right panels). In agreement with what can be seen from Fig.~\ref{fig:scale_height_density_maximum}, the oscillating dust mid-plane layer extends to greater heights in the latter scenario than in the former one. In addition, the dust is more uniformly distributed in the radial dimension in the former scenario. Interestingly, the region between~\mbox{$r=15~\au$} and~$30~\au$ is depleted of dust in both simulations of the scenario \textit{SIafterVSI} that are presented in the figure. The reason for this depletion is unclear.

The time- and radius-dependence of the dust-to-gas surface density ratio in the scenario \textit{SIwhileVSI} is depicted in the lower left panel of the figure. The streaming instability causes the dust to radially accumulate in overdensities. Comparable, azimuthally elongated filaments are found in three-dimensional simulations of the instability \citep{Johansen2007b, Bai2010b, Kowalik2013}. As noted above, the dust concentration in this scenario is sufficiently strong for the streaming instability to induce planetesimal formation.

From the lower middle and right panels, it can be seen that similar, but significantly denser dust concentrations form in the scenario \textit{SIafterVSI}. If the surface density ratio is higher, that is to say the total dust mass is greater, more dust is comprised in overdensities. The radial drift of overdensities is reduced in both scenarios. Only in the scenario \textit{SIafterVSI}, though, some of the accumulations are dense enough for their radial drift to be nearly completely halted.

\citet{Stoll2016} perform simulations of the vertical shear instability in which dust and the drag exerted on it by the gas are included. They find that pressure fluctuations which are induced by the vertical shear instability lead to a radial concentration of the dust. This concentration is strongest if the dimensionless stopping time of the dust is close to unity. Nevertheless, because they do not take the drag back-reaction of the dust onto the gas into account, the radial drift speed does not depend on the dust density in their simulations.

In the vicinity of the dust overdensities in the scenario \textit{SIafterVSI}, we find the streaming instability to contribute to the driving of turbulence. At the end of the simulation of the scenario \textit{SIafterVSI} with a surface density ratio of~$2\%$ --~$55~\kyr$ after the beginning of the simulation and~$5~\kyr$ after the initialization of the dust -- several overdensities can be found between~\mbox{$r=30~\au$} and~$40~\au$ as well as around~$60~\au$ (see the lower middle panel of Fig.~\ref{fig:density}). From the right panel of Fig.~\ref{fig:Mach_number_dust_outer_region}, it is evident that at this time and these radii small-scale perturbations are present. We associate these perturbations with the streaming instability because they are not observable in our model of the vertical shear instability only.

Thus, it is most probably a combination of the vertical shear instability and the streaming instability that induces the dust accumulation in this scenario. We speculate that the vertical shear instability first concentrates the dust in weak overdensities, as shown by \citet{Stoll2016}. Then, the streaming instability causes the growth of these seeds to the dense accumulations that we find in our simulations.

The fact that the dust is radially more concentrated in the scenario \textit{SIafterVSI} despite its vertical diffusion being stronger in this scenario is consistent with the results of the numerical study by \citet{Yang2018}. They show that the radial dust concentration owing to the streaming instability is not significantly affected by the vertical dust diffusion that is induced by non-ideal MHD turbulence. 

\section{Global simulations of the streaming instability}
\label{sect:SI}
To study the vertical and radial gas motions which are induced by the streaming instability as well as the dust scale height, we employ our model including dust and locally adiabatic gas. The latter entails that the vertical shear instability is quenched by vertical buoyancy.

\begin{figure*}[t]
\centering
\begin{minipage}{0.49\textwidth}
\centering
\includegraphics[width=\textwidth]{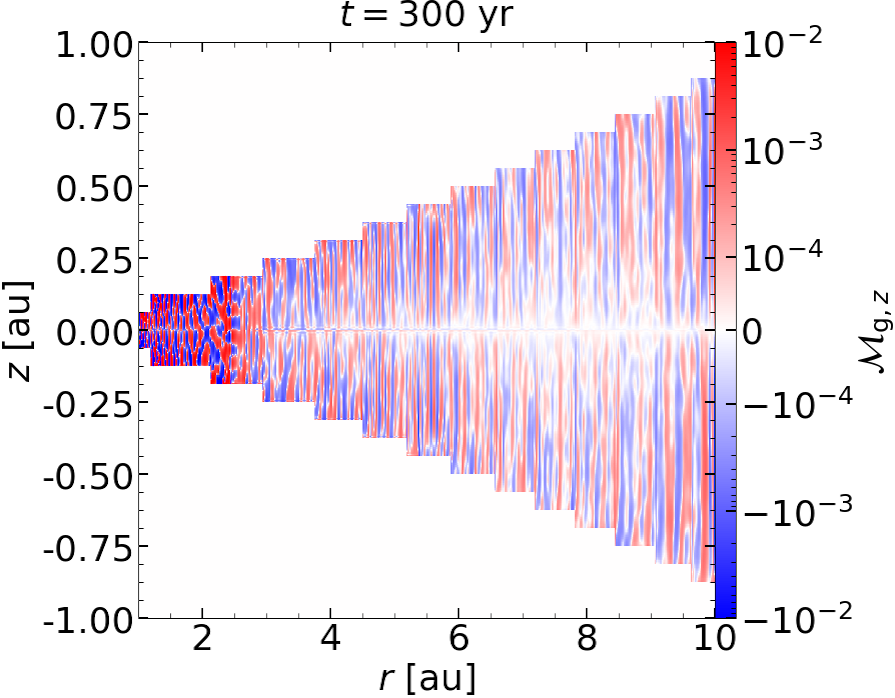}
\end{minipage}
\hfill
\begin{minipage}{0.49\textwidth}
\centering
\includegraphics[width=\textwidth]{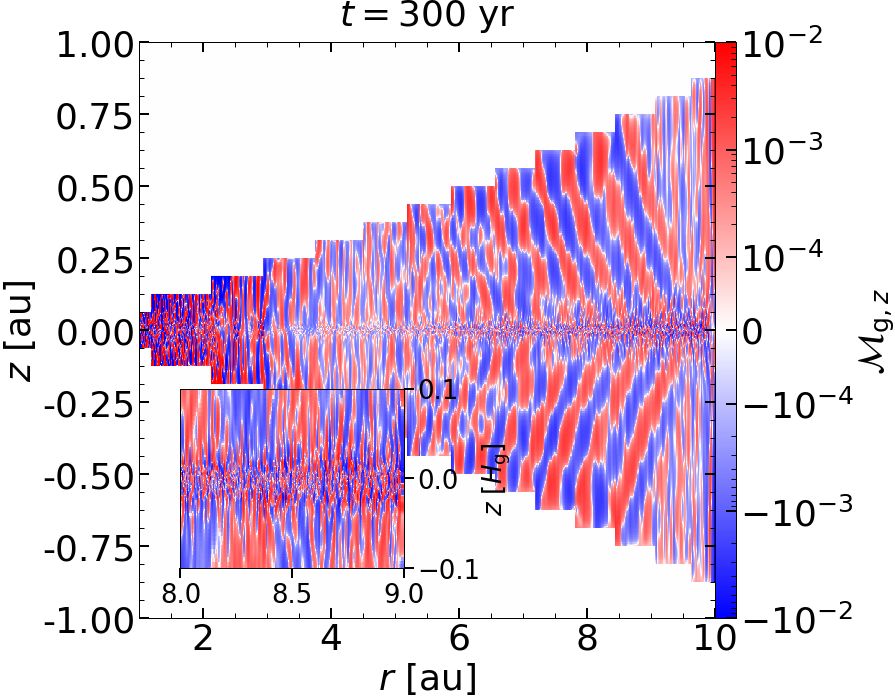}
\end{minipage}
\caption{Mach number of the vertical gas velocity as a function of~$r$ and~$z$ in simulations of dust and a locally adiabatic gas. The Mach number at radii less than~${\sim}3~\au$ is similarly high in the simulation without (\textit{adi\_Lr=9au}; left panel) and the one with dust (\textit{adi\_Z=0.02\_Lr=9au}; right panel). This is in spite of the streaming instability operating in the latter simulation, but not in the former one. At larger radii in the simulation including dust, the streaming instability induces large-scale perturbations with~\mbox{$\mathcal{M}_{\rm g,z}\approx10^{-3}$} away from the mid-plane. These perturbations are similar in shape to the ones which the vertical shear instability causes. Yet, they are weaker, are bent inwards rather than outwards and are not symmetric with respect to the mid-plane (compare with Fig.~\ref{fig:Mach_number_no_dust_outer_region}). In the inlay, which extends to~$0.1$ gas scale heights~$H_{\rm g}$ above and below mid-plane, small-scale perturbations with~\mbox{$\mathcal{M}_{\rm g,z}\approx10^{-2}$} in the mid-plane can be seen.}
\label{fig:Mach_number_inner_region}
\end{figure*}

The Mach number of the vertical motions in this model as well as its dust-free equivalent are shown in Fig.~\ref{fig:Mach_number_inner_region}. Since in the latter model (left panel) the streaming instability is not active, the turbulence in this model is most probably a numerical artifact which is caused by vertical shear that results from imperfect boundary conditions. At radii greater than~${\sim}3~\au$, this artificial turbulence is negligibly weak compared to the turbulence which is driven by the streaming instability in the model with dust (right panel). At smaller radii, however, we can not distinguish between artificial turbulence and turbulence that is caused by the streaming instability. We therefore exclude these radii from the following analysis.

In the model including dust, the streaming instability induces small-scale perturbations with a turbulent strength of~\mbox{$\mathcal{M}_{\rm g,z}\approx0.01$} in the dust mid-plane layer (see the inlay in the right panel) and somewhat weaker large-scale perturbations away from it. The small-scale perturbations are similar to the ones which the streaming instability gives rise to in the model in which both it and the vertical shear instability operate (see Fig.~\ref{fig:Mach_number_dust_outer_region}). We find both kinds of perturbations to be largely isotropic (see also Fig.~\ref{fig:Mach_number_ratio}).

The large-scale perturbations resemble the perturbations that are caused by the vertical shear instability in that their radial-to-vertical wavelength ratio is much less than one (see Fig.~\ref{fig:Mach_number_no_dust_outer_region}). In contrast to these, their symmetry is odd with respect to the mid-plane and they are bent inwards. This bending of the perturbations can be explained by gas moving both vertically away from the mid-plane and radially outwards, the latter because of the inward radial drift of the dust and the conservation of angular momentum.

In the local shearing box simulations that are presented by \citet{Li2018}, the streaming instability gives rise to the same two kinds of perturbations with a similar turbulent strength (compare with their Fig.~2). The large-scale perturbations which they observe are bent outwards rather than inwards, though. Nevertheless, these perturbations are likely suppressed by the stronger ones which the vertical shear instability gives rise to under more realistic conditions, i.e., in simulations including heating and cooling by radiation rather than an adiabatic equation of state \citep{Stoll2014, Flock2017}.

\subsection{Vertical gas velocity and dust scale height}
\begin{figure*}[t]
\centering
\begin{minipage}{0.49\textwidth}
\centering
\includegraphics[width=\textwidth]{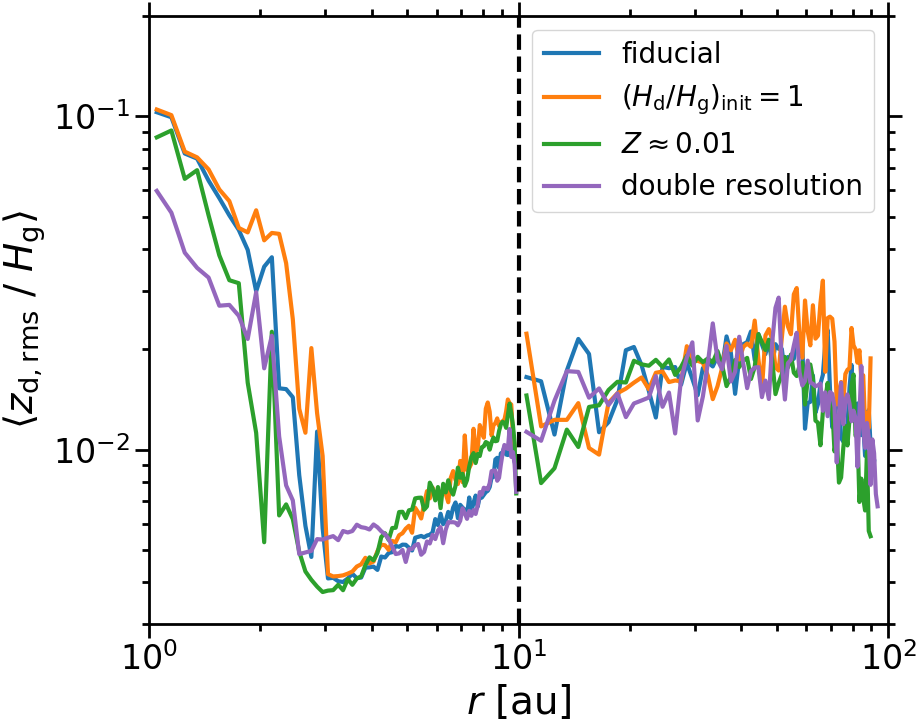}
\end{minipage}
\hfill
\begin{minipage}{0.49\textwidth}
\centering
\includegraphics[width=\textwidth]{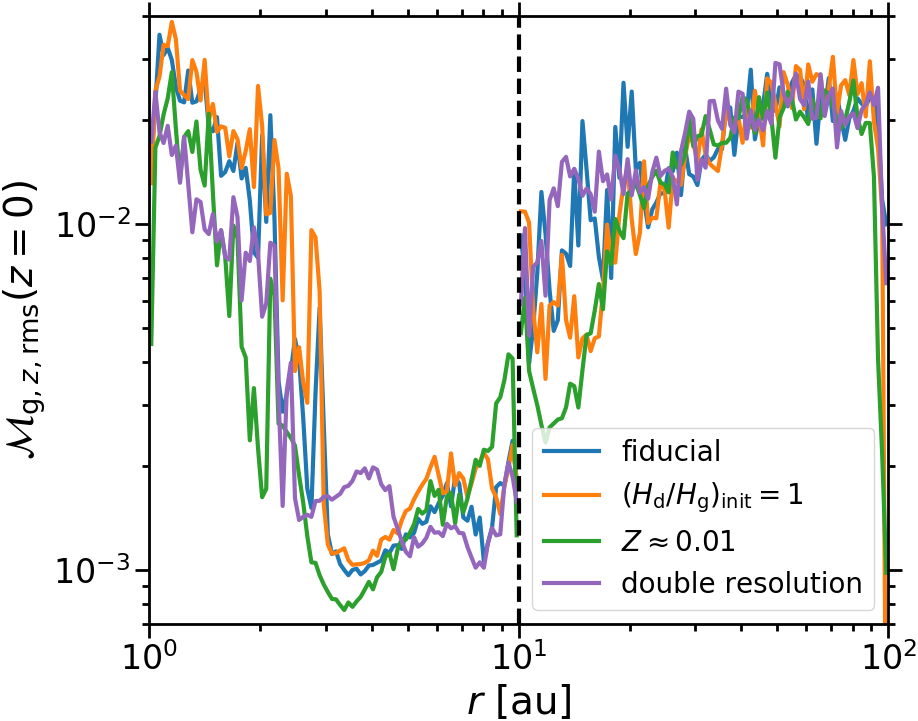}
\end{minipage}
\caption{Ratio of the dust scale height, as the RMS of~$z_{\rm d}$, to the gas scale height~$H_{\rm g}$ (left panel) as well as mass-weighted RMS of~$\mathcal{M}_{{\rm g},z}$ in the mid-plane (right panel). Both quantities are depicted as functions of~$r$. To compute the RMS values, we average over~$50~\yr$ or~$500~\yr$ after the dust scale height has reached an equilibrium value in the simulation domains spanning~\mbox{$1~\au\leq r\leq 10~\au$} or~\mbox{$10~\au\leq r\leq 100~\au$}, respectively. The dashed line marks the boundary between these domains. The model with an initial dust-to-gas scale height ratio of~\mbox{$(H_{\rm d}/H_{\rm g})_{\rm init}=0.1$}, a dust-to-gas surface density ratio of~\mbox{$Z=2\%$}, and the fiducial resolution (blue line) is shown together with the models that deviate from this fiducial one in that~\mbox{$(H_{\rm d}/H_{\rm g})_{\rm init}=1$}, that~\mbox{$Z=1\%$} (green line), or that the initial and maximum resolution are doubled (purple line). While both the dust scale height and the Mach number depend on the radius, they are largely independent of the initial dust scale height, the surface density ratio, and the resolution.}
\label{fig:scale_height_vertical_motions}
\end{figure*}

In Fig.~\ref{fig:scale_height_vertical_motions}, we show the dust-to-gas scale height ratio as well as the Mach number of the vertical gas velocity in the mid-plane after dust settling and vertical diffusion have reached an equilibrium. We find both the dust scale height and the turbulent strength to be similar in the model with a dust-to-gas surface density ratio of~$2\%$ and the model with a ratio of~$1\%$. 
%This is in spite of the dust concentration owing to the streaming instability being sufficiently strong to lead to planetesimal formation in the former case, but not in the latter one \citep{Carrera2015, Yang2017}. 
The equilibrium scale height and the Mach number do not depend significantly on the initial dust scale height either, regardless of the dust taking longer to sediment to the mid-plane if it is greater. Furthermore, our results are converged with respect to resolution.

\citet{Bai2010b} study local shearing box simulations of the streaming instability with surface density ratios of~$1\%$,~$2\%$, or~$3\%$. They derive the strength of the vertical diffusion as well as the scale height of the dust from fitting a Gaussian distribution to the vertical dust density profile. In contrast to us, they find the vertical diffusion to be stronger and the scale height to increase with the dust-to-gas surface density ratio if it is less than a threshold value. Above this threshold value, which depends on the particle size, the turbulent strength decreases again, and the dust settles to smaller scale heights.

The dust scale height is equal to~${\sim}1\%$ of the gas scale height at all radii in our model. The streaming instability gives rise to a similar dust scale height in our model in which both it and the vertical shear instability are active and begin to grow at the same time (see the left panel of Fig.~\ref{fig:scale_height_density_maximum}). Comparable dust scale heights are further found in local shearing box simulations, e.g., the ones that were conducted by \citet{Yang2014} and \citet{Carrera2015}. For these scale heights and the dust-to-gas surface density ratios of the order of~$1\%$ which we simulate, the ratio of the dust to the gas density in the mid-plane is of order unity, and the linear growth rate of the instability is largest \citep{Youdin2005}.

The thickness of the dust layer is set in a self-regulatory manner if the turbulent diffusion of the dust is caused by the streaming instability \citep{Bai2010b}: If the dust scale height is greater than the equilibrium value, the instability induces weaker turbulence, which does not balance the dust settling towards mid-plane. On the other hand, if the scale height is less than the equilibrium value, the instability drives overly strong turbulence, which lifts the dust away from the mid-plane.

At radii~\mbox{$r\gtrsim$10~\au}, the Mach number is consistent with the observed values of the order of~$10^{-2}$ \citep{Flaherty2015, Flaherty2017, Flaherty2018, Pinte2016, Ohashi2019}. The turbulent strength is comparable in our model in which both the vertical shear instability and the streaming instability operate, but the latter drives the turbulence in the dust layer. \citet{Pinte2016} and \citet{Ohashi2019} derive the turbulent strength from the dust scale height, which is regulated by the streaming instability if it is the main source of turbulence in the dust layer of protoplanetary disks. We note that these authors observe a dust-to-gas scale height ratio of~$10\%$ and one third, respectively, which is an order of magnitude larger than the ratio we find. This can be explained by them considering micron- and millimeter-sized dust grains, which are elevated to greater heights than the cm-sized ones in the simulations which we present in Fig.~\ref{fig:scale_height_vertical_motions}.

\subsection{Dependence of turbulent strength on dust stopping time and radial gas pressure gradient}
\begin{figure}[t]
\centering
\includegraphics[width=\columnwidth]{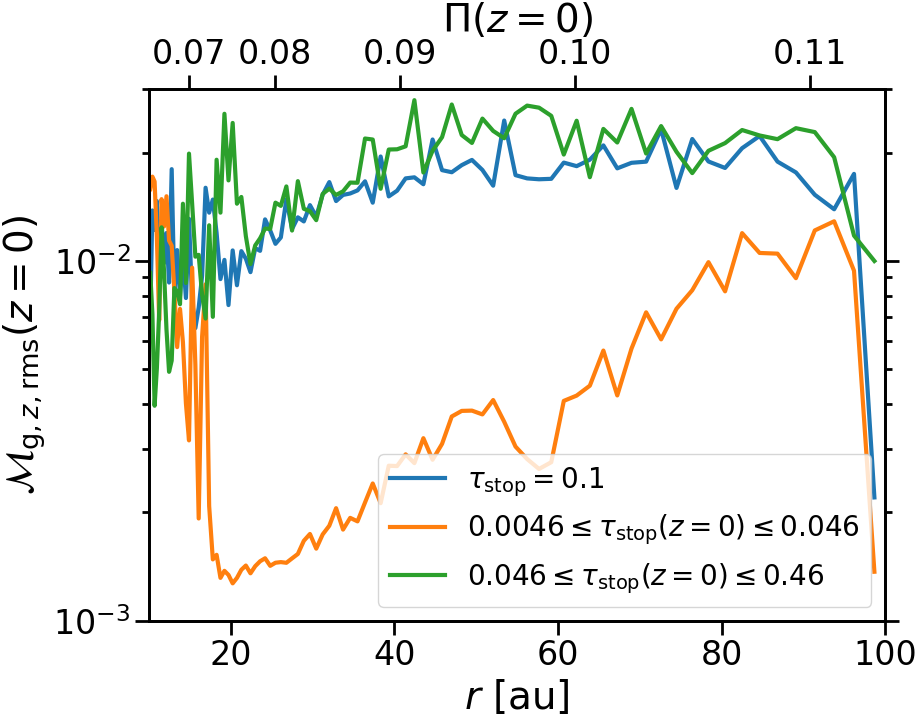}
\caption{RMS of~$\mathcal{M}_{{\rm g},z}$ in the mid-plane as a function of~$r$ (lower abscissa) and the dimensionless gas pressure gradient parameter~$\Pi$ in the mid-plane (upper abscissa). The RMS values are calculated using the mass-weighted mean over~$500~\yr$ after an equilibrium value of the dust scale height is reached. In the simulation \textit{adi\_Z=0.02\_Lr=90au\_taustop=0.1} (blue line), in which the dimensionless dust stopping time~$\tau_{\rm stop}$ is fixed at~$0.1$, the Mach number increases with the pressure gradient parameter if~$\Pi\lesssim0.09$, but is approximately constant if the pressure gradient is stronger. The Mach number in this one and the simulation \textit{adi\_Z=0.02\_Lr=90au} (green line), in which the stopping time increases from~$0.046$ to~$0.46$, is comparable at all radii. That is, the turbulent strength does not depend significantly on the stopping time in this stopping time regime. The Mach number is overall lower, but depends more strongly on the radius in the simulation \textit{adi\_Z=0.02\_Lr=90au\_a=3mm} (orange line), in which the stopping time ranges from~$0.0046$ to~$0.046$. In this regime, the turbulent strength increases with both the strength of the pressure gradient and the stopping time.}
\label{fig:vertical_motions_pressure_gradient}
\end{figure}

We find that the strength of the turbulence that is driven by the streaming instability, like the linear growth rate of the instability \citep{Youdin2005}, increases with the speed of the radial dust drift. The drift is faster if the dimensionless stopping time of the dust is closer to one and the radial gas pressure gradient is stronger. Both quantities increase with the radius in our model of dust with a fixed size (see Eqs.~\ref{eq:pressure_gradient} and~\ref{eq:stopping_time}). From Fig.~\ref{fig:scale_height_vertical_motions}, it is evident that both the Mach number and the dust scale height consequently as well are greater at larger radii.

To analyze how strongly the turbulent strength depends on the stopping time and the magnitude of the pressure gradient individually, we compare three simulations with different dimensionless stopping times: in one, it is fixed at~$0.1$, while in the other two it ranges from~$0.0046$ to~$0.046$ and from~$0.046$ to~$0.46$, respectively. The Mach number in each of these simulations is depicted in Fig.~\ref{fig:vertical_motions_pressure_gradient} as a function of the radius and the pressure gradient strength. We find an equivalent radius-dependence of the dust scale height in the three simulations.

The turbulent strength increases with both the dimensionless pressure gradient parameter and the dimensionless stopping time, until it saturates for respective values of~\mbox{$\Pi\approx0.1$} and of~\mbox{$\tau_{\rm stop}\approx0.05$}. This is consistent with \citet{Bai2010a} and \citet{Bai2010b} finding the scale height of the dust and its vertical diffusion to increase with the strength of the pressure gradient and the stopping time. The dependence on the pressure gradient strength is evident when considering only the simulation with a fixed stopping time (blue line in the figure).

At all radii in the simulation with a dimensionless stopping time of~$0.1$ and the simulation with stopping times greater than~$0.046$ (green line), the turbulent strength is about the same. Thus, it is largely independent of the stopping time if~\mbox{$\tau_{\rm stop}\gtrsim0.05$}. In contrast, in the simulation with stopping times less than~$0.046$ (orange line) the turbulence is overall weaker, but its strength increases more strongly with the radius than in the other two simulations. That is, if~\mbox{$\tau_{\rm stop}\lesssim0.05$} the strength depends not only on the pressure gradient magnitude, but also on the stopping time.

\subsection{Radial gas velocity}
\begin{figure}[t]
\centering
\includegraphics[width=\columnwidth]{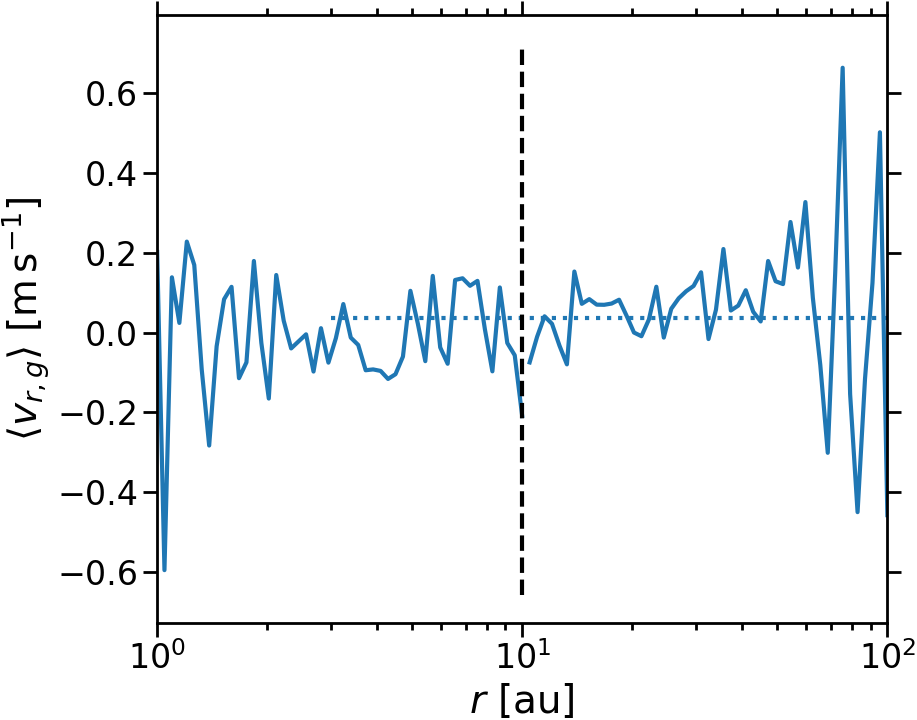}
\caption{Radial gas velocity (solid line) as a function of~$r$ in the simulations \textit{adi\_Z=0.02\_Lr=9au} and \textit{adi\_Z=0.02\_Lr=90au}. (The dashed lines marks the boundary between the domains of the two simulations.) The velocity is computed as the mass-weighted average over the vertical domain extent and a time span of $50~\yr$ and $500~\yr$, respectively, after an equilibrium dust scale height has been attained in the former and the latter simulation. The average radial velocity at~$r\geq 3~\au$ is plotted as a dotted line and amounts to~$0.035~\m\,\s^{-1}$. This outward motion is caused by the dust drifting radially inwards and its angular momentum being transferred to the gas. It is evident, though, that the mean of the velocity is less than its standard deviation, which is equal to~$0.16~\m\,\s^{-1}$. This is a result of the turbulent motions that the streaming instability gives rise to.}
\label{fig:radial_motions}
\end{figure}

In Fig.~\ref{fig:radial_motions}, we show the radial velocity of the gas. The gas on average moves outward as a consequence of the inward radial drift of the dust and, as the total angular momentum is conserved, the transfer of angular momentum from the dust to the gas. Yet, the turbulent motions that are induced by the streaming instability lead to the variance in the gas velocity being greater than the speed of this mean outward motion. That is, the streaming instability does not cause a sustained inward transport of gas mass (and outward transport of angular momentum) that contributes to the observed stellar accretion \citep{Alcala2017}.

\section{Discussion}
\label{sect:discussion}
\subsection{Implications for turbulence and planetesimal formation in protoplanetary disks}
We have studied the interaction of two instabilities, the vertical shear instability and the streaming instability. Both instabilities appear to be robust: To operate, the vertical shear instability requires a vertical rotation profile and a sufficiently short gas cooling time scale. \citet{Lin2015} show that in the MMSN model, the instability can grow between~$r\approx5~\au$ and~$50~\au$. On the other hand, only dust and a radial gas pressure gradient are necessary for the streaming instability to be active.

We have considered two scenarios: In the first scenario, the vertical shear instability and the streaming instability start to grow at the same time. In this scenario, the turbulence in the dust layer is driven by the streaming instability. If the vertical shear instability has already saturated beforehand, as is the case in the second scenario, it remains the main source of turbulence even while the streaming instability is active. 

It is unclear which of these scenarios is more realistic: Monte-Carlo simulations of dust coagulation show that \mbox{(sub-)}micron-sized grains grow to millimeter- or centimeter-sized aggregates within~$10^3$ to~$10^4$ orbital periods \citep{Zsom2010, Lorek2018}. However, it is uncertain at which point during the formation of a protoplanetary disk dust growth commences as well as when the conditions in a disk become conducive to the growth of the vertical shear instability and the streaming instability.

The vertical shear instability gives rise to stronger turbulence, and thus a stronger vertical dust diffusion, than the streaming instability. Nonetheless, we find this instability, in combination with the streaming instability, to cause the dust to be substantially more concentrated in the radial dimension than the streaming instability alone. We note that previous studies show the dust accumulation owing to the streaming instability to be sufficient to lead to planetesimal formation for the dust-to-gas surface density ratios we consider.

That is, for a given dust size the critical surface density ratio that is required for planetesimal formation may be significantly lower if not only the streaming instability, but the vertical shear instability and the streaming instability together induce the dust concentration. The critical value exceeds~$1\%$ for all dust sizes in the former case \citep{Carrera2015, Yang2017}, but might be lower than this canonical interstellar medium value in the latter case. 

\subsection{Limitations of our numerical study}
Both the linear vertical shear instability \citep{Nelson2013, Barker2015} and the linear streaming instability \citep{Youdin2005} are axisymmetric in nature. Nonetheless, slight deviations from this symmetry are found in simulations of the non-linear regime of both the former instability \citep{Nelson2013, Stoll2014} and the latter one \citep{Kowalik2013}. These deviations are not captured by our two-dimensional simulations. 

\citet{Umurhan2019} analytically examine the linear streaming instability in connection with the $\alpha$-model for protoplanetary disk turbulence \citep{Shakura1973}. They find that turbulence reduces the growth rate of the instability compared to the purely laminar case. This analysis is not applicable if the streaming instability itself is the dominant source of turbulence, however.

Similarly, \citet{Gole2020} study planetesimal formation owing to the streaming instability in local shearing box simulations with driven Kolmogorov-like turbulence that is not affected by the presence of the dust. In their simulations, planetesimals formation is hampered by turbulence and quenched if the turbulent Mach number is greater than~${\sim}10^{-2}$.

We have studied only models with a single dust size (or a single dimensionless dust stopping time). \citet{Krapp2019} show that the linear growth rate of the streaming instability can decrease if multiple dust species with a distribution of sizes rather than a single species are simulated. Yet, this effect has not been found in simulations of the non-linear streaming instability \citep{Bai2010b, Schaffer2018}. 

Furthermore, we have not taken into account other hydrodynamic instabilities like the convective overstability \citep{Klahr2014, Lyra2014}, the subcritical baroclinic instability \citep{Klahr2003, Klahr2004, Lyra2014}, or the zombie vortex instability \citep{Marcus2015, Marcus2016, Lesur2016}. These instabilities do not operate in our model because the former two require finite gas cooling time scales -- we simulate either a locally adiabatic or a locally isothermal gas -- while the latter is three-dimensional.
%The streaming instability dominates over the Kelvin-Helmholtz instability that results from the vertical shear between gas layers with different dust-to-gas density ratios and thus orbital velocities \citep{Bai2010b}.

In addition, our simulations do not include magnetic fields. \citet{Cui2019} study the vertical shear instability in non-ideal MHD simulations. They find that magnetized disk winds and the vertical shear instability can coexist. Under conditions that are typical of protoplanetary disks, the instability induces a comparable turbulent strength in their simulations and in purely hydrodynamical ones like ours. Nonetheless, if the magnetization is enhanced or the gas is more strongly coupled to the magnetic field, the instability causes weaker turbulence.

\section{Summary}
\label{sect:summary}
We have presented two-dimensional axisymmetric global numerical simulations of protoplanetary disks spanning orbital radii between~$1~\au$ and~$100~\au$. The simulations include Lagrangian particles to model the dust, the mutual drag between dust and gas, as well as the radial and vertical stellar gravity. We have used the FLASH Code to conduct these simulations, which has allowed us to apply adaptive mesh refinement to increase the resolution locally in and close to the dust layer in the mid-plane of the disks.

Employing these simulations, we have investigated the turbulence that is driven by the vertical shear instability and the streaming instability individually as well as the interaction of the two instabilities. The results of our study can be summarized as follows:
\begin{itemize}
\item We have conducted simulations of the vertical shear instability only with vertical domain extents of one or two gas scale heights above and below the mid-plane. Only the latter vertical size is sufficient to reproduce the turbulent strength that is found in previous numerical studies of vertically larger domains. The Mach number of the vertical gas motions is of the order of~$10^{-1}$ in the saturated state of the instability \citep{Flock2017}.
\item[] 
\item If both the vertical shear instability and the streaming instability start to grow simultaneously, we find the turbulence in the dust mid-plane layer to be mainly driven by the streaming instability. This is most likely the result of a combination of two effects: In the mid-plane, the streaming instability grows faster in turbulent strength than the vertical shear instability. Furthermore, the weight of the dust induces an effective buoyancy in the gas that quenches the vertical shear instability \citep{Lin2017, Lin2019}.
\item[]
\item The vertical dust settling and the turbulent diffusion that is induced by the streaming instability attain an equilibrium if the dust-to-gas scale height ratio is equal to~${\sim}1\%$. 
%This value is consistent with the ones found in local shearing box simulations of the streaming instability \citep{Yang2014, Carrera2015}. 
The dust scale height is set in a self-regulatory way if the streaming instability gives rise to the diffusion of the dust \citep{Bai2010b}: If the scale height is less than the equilibrium value, then the turbulent strength is greater than the equilibrium strength, and the dust is lifted away from the mid-plane.
\item[]
\item We show that the streaming instability drives isotropic turbulence with a Mach number of up to~${\sim}10^{-2}$. This is in agreement with observed values in protoplanetary disks \citep{Flaherty2015, Flaherty2017, Flaherty2018, Pinte2016, Ohashi2019}. In particular, \citet{Pinte2016} and \citet{Ohashi2019} obtain this Mach number from the scale height of the dust disks surrounding HL Tau and HD 163296, respectively. That is, they would probe the turbulent strength that is induced by the streaming instability if this instability were the primary source of turbulence in the dust layer of protoplanetary disks.
\item[]
\item Both the equilibrium dust scale height and the Mach number that are induced by the streaming instability are largely independent of the dust-to-gas surface density ratio and the initial dust scale height.
% in our model. \citet{Bai2010b} conduct local shearing box simulations with the same surface density ratios as we study, but find the vertical dust diffusion to depend on the surface density ratio.
%\item
The turbulent strength, and with it the scale height, increases with the speed of the radial dust drift. That is, the turbulence is stronger if the dust stopping time of the radial gas pressure gradient is greater. The strength saturates for dimensionless stopping times of~${\sim}0.05$ and dimensionless pressure gradient parameters, as defined by \citet{Bai2010b}, of~${\sim}0.1$.
%\citet{Bai2010a} and \citet{Bai2010b} as well show that the strength of the vertical diffusion and the scale height of the dust is greater for larger stopping times and stronger pressure gradients.
\item[]
\item In contrast, if the vertical shear instability has attained a saturated state before we introduce the dust into our simulations, then this instability remains the primary source of turbulence in the dust layer. It gives rise to stronger turbulence than the streaming instability, which elevates the dust to greater scale heights. For a dust-to-gas surface density ratio of~$2\%$, the instability induces a Mach number of~${\sim}10^{-1}$ and a dust scale height of~${\sim}10\%$ of the gas scale height. Nevertheless, if the surface density ratio is higher, the instability is more strongly quenched by the dust-induced buoyancy. 
\item[]
\item We find that a combination of the vertical shear instability and the streaming instability leads to a considerably stronger radial concentration of the dust than the streaming instability only. The dust accumulations are dense enough for their radial drift to be halted almost completely. This is despite the vertical shear instability inducing stronger vertical diffusion than the streaming instability. We speculate that the vertical shear instability induces the formation of weak overdensities that seed the streaming instability. The streaming instability in turn causes strong dust concentration that would likely lead to planetesimal formation in simulations including the self-gravity of the dust.
%\citet{Stoll2016} show that in their simulations, in which the drag that is exerted by the dust on the gas is not included, the dust is concentrated in the radial dimension owing to pressure fluctuations which are caused by the vertical shear instability. 
\end{itemize}

\begin{acknowledgements}
We thank the anonymous referee and Chao-Chin Yang for their constructive feedback that helped to improve this paper. To analyze and visualize the simulations, the Python packages yt\footnote{\url{http://yt-project.org}} \citep{Turk2011}, Matplotlib\footnote{\url{https://matplotlib.org}} \citep{Hunter2007}, and NumPy\footnote{\url{https://numpy.org}} \citep{Oliphant2006} have been used. The FLASH Code has in part been developed by the DOE NNSA-ASC OASCR Flash Center at the University of Chicago. Computational resources employed to conduct the simulations presented in this paper were provided by the Regionales Rechenzentrum at the University of Hamburg, by the Norddeutscher Verbund für Hoch- und Höchstleistungsrechnen (HLRN), and by the Swedish Infrastructure for Computing (SNIC) at LUNARC at Lund University. U.S. thanks the University of Hamburg for granting him a scholarship to fund his doctoral studies. A.J. is thankful for research support by the European Research Council (ERC Consolidator Grant 724687-PLANETESYS), the Knut and Alice Wallenberg Foundation (Wallenberg Academy Fellow Grant 2017.0287), and the Swedish Research Council (Project Grant 2018-04867). U.S. and R.B. gratefully acknowledge financial support by the Deutsche Forschungsgemeinschaft (DFG), grant BA 3706/18-1. R.B. is thankful for support by the Excellence Cluster 2121 ``Quantum Universe'' which is funded by the DFG.
\end{acknowledgements}

\bibliography{Paper} 

\begin{appendix}

\section{Leapfrog algorithm for cylindrical geometries}
\label{sect:Leapfrog_algorithm}
\subsection{Implementation}
We implement a second-order accurate, explicit Leapfrog algorithm for the time integration of particles in cylindrical geometries. The algorithm is based on the one that has been developed by \citet{Boris1970} for charged particles in simulations including electric and magnetic fields. It has been described as well by, e.g., \citet{Delzanno2013}.

For the update of the vertical components of the particle velocity and position, we adopt the Leapfrog algorithm for Cartesian geometries that is part of the FLASH Code:
\begin{equation}
v_{z,i}^{1/2}=v_{z,i}^0+\frac{1}{2}a_{z,i}^0\Delta t^0,
\end{equation}
\begin{equation}
v_{z,i}^{n+1/2}=v_{z,i}^{n-1/2}+A_n a_{z,i}^n+B_n a_{z,i}^{n-1},\text{ and}
\end{equation}
\begin{equation}
z_i^{n+1}=z_i^n+v_{z,i}^{n+1/2}\Delta t^n,
\end{equation}
where~$z_i^n$,~$v_{z,i}^n$, and~$a_{z,i}^n$ are the vertical position, velocity, and acceleration of the~$i$-th particle at the~$n$-th time step~$\Delta t^n$, respectively. The acceleration is computed employing cloud-in-cell mapping between the grid and the particles. In our simulations, it is due to the stellar gravity only. We explain in Appendix~\ref{sect:drag} how we take account of the acceleration that is caused by the drag of the gas onto the particles. The coefficients~$A_n$ and~$B_n$ are given by
\begin{equation}
A_n=\frac{1}{2}\Delta t^n+\frac{1}{3}\Delta t^{n-1}+\frac{1}{6}\frac{\Delta t^{n^2}}{\Delta t^{n-1}}\text{ and}
\end{equation}
\begin{equation}
B_n=\frac{1}{6}\left(\Delta t^{n-1}-\frac{\Delta t^{n^2}}{\Delta t^{n-1}}\right).
\end{equation}

The radial and azimuthal velocity and position components are advanced as in Cartesian coordinates, followed by a transformation from these to cylindrical coordinates:
\begin{enumerate}
  \item The velocity is calculated as in a Cartesian geometry, i.e., inertial forces are disregarded:
\begin{equation}
v_{\{r,\phi\},i}^{1/2'}=v_{\{r,\phi\},i}^0+\frac{1}{2}a_{\{r,\phi\},i}^0\Delta t^0 \text{ and}
\end{equation}
\begin{equation}
v_{\{r,\phi\},i}^{n+1/2'}=v_{\{r,\phi\},i}^{n-1/2}+A_n a_{\{r,\phi\},i}^n+B_n a_{\{r,\phi\},i}^{n-1},
\end{equation}
where the subscript~$\{r,\phi\}$ denotes the radial or azimuthal component.
  \item The position is updated in Cartesian geometry and then transformed from Cartesian to cylindrical geometry:  
\begin{equation}
x_i^{n+1}=r_i^n+v_{r,i}^{n+1/2'}\Delta t^n,
\end{equation}
\begin{equation}
y_i^{n+1}=v_{\phi,i}^{n+1/2'}\Delta t^n,
\end{equation}
\begin{equation}
r_i^{n+1}=\sqrt{x_i^{n+1^2}+y_i^{n+1^2}},\text{ and}
\end{equation}
\begin{equation}
\phi_i^{n+1}=\phi_i^n+\alpha_i^{n+1},
\end{equation}
where~$r_i^n$ and~$\phi_i^n$ are the radial and azimuthal position, respectively, and the angle~$\alpha_i^{n+1}$ can be computed as
\begin{equation}
\alpha_i^{n+1}=\arccos\left({\frac{x_i^{n+1}}{r_i^{n+1}}}\right)=\arcsin\left({\frac{y_i^{n+1}}{r_i^{n+1}}}\right)
\end{equation}
  \item The velocity is corrected to reflect the transformation from Cartesian to cylindrical geometry, which entails the consideration of inertial forces:
\begin{equation}
v_{r,i}^{n+1/2}=\cos\left(\alpha_i^{n+1}\right)v_{r,i}^{n+1/2'}+\sin\left(\alpha_i^{n+1}\right)v_{\phi,i}^{n+1/2'}\text{ and}
\end{equation}
\begin{equation}
v_{\phi,i}^{n+1/2}=-\sin\left(\alpha_i^{n+1}\right)v_{r,i}^{n+1/2'}+\cos\left(\alpha_i^{n+1}\right)v_{\phi,i}^{n+1/2'}.
\end{equation}
\end{enumerate} 

The velocity at a full time step can be computed from the one at a half time step as follows:
\begin{equation}
\bm{v}_i^n=\bm{v}_i^{n-1/2}+\frac{1}{2}\left(A_n\bm{a}_i^n+B_n\bm{a}_i^{n-1}\right)
\label{eq:full-step_velocity}
\end{equation}
We note that this full-step velocity is second-order accurate, while the error in the half-step velocity is proportional to the time step.

\subsection{Test}
We test our implementation by using the Leapfrog algorithm for Cartesian geometries that is already included in the FLASH Code as a benchmark: We conduct two analogous simulations of a particle in the gravitational potential of a point mass, one with a two-dimensional cylindrical and one with a three-dimensional Cartesian geometry.

The initial position of the particle, relative to the position of the point mass, can be expressed as~\mbox{$(r,z)=(3~\au,-1~\au)$} in cylindrical and by~\mbox{$(x,y,z)=(3~\au,0~\au,-1~\au)$} in Cartesian coordinates. To establish an initial balance between the centrifugal and the radial gravitational force which are exerted on the particle, we set its velocity to
\begin{equation}
\begin{cases}
v_{\phi}=\sqrt{\frac{GM}{\left(r^2+z^2\right)^{3/2}}r^2}&\text{(cylindrical geometry) or}\\
v_y=\sqrt{\frac{GM}{\left(x^2+y^2+z^2\right)^{3/2}}\left(x^2+y^2\right)}&\text{(Cartesian geometry),}\\
\end{cases}
\end{equation}
where~\mbox{$M=1~M_{\sun}$} is the point mass. Cloud-in-cell mapping is applied to calculate the gravitational acceleration of the particle. The grid cell edge length is fixed at~$0.025~\au$.

\begin{figure}[t]
\centering
\includegraphics[width=\columnwidth]{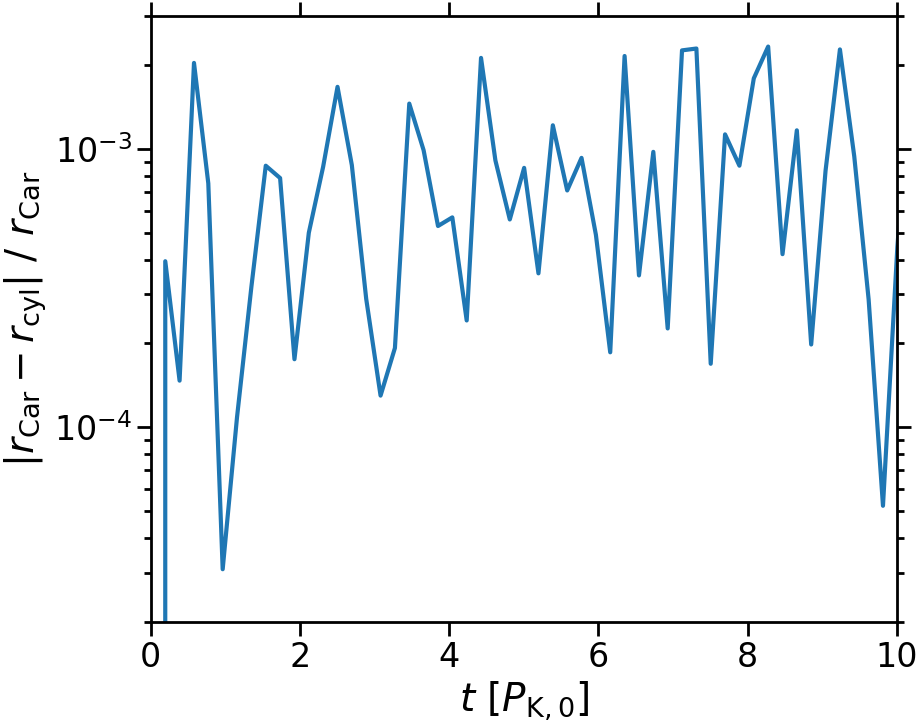}
\caption{Error of the radial coordinate $r_{\rm cyl}$ that is computed by our implementation of the Leapfrog algorithm for cylindrical geometries relative to the coordinate~\mbox{$r_{\rm Car}=\sqrt{x^2+y^2}$} that is calculated by the already implemented algorithm for Cartesian geometries. We conduct two equivalent simulations of a particle in a point mass gravitational potential, one using the former and one applying the latter algorithm. Both simulations end after ten Keplerian orbital periods at the initial position of the particle~$P_{\rm K,0}$. As can be seen, the relative error, while varying, is at most~$2\times10^{-3}$.}
\label{fig:Leapfrog_test}
\end{figure}

In Fig.~\ref{fig:Leapfrog_test}, we show the relative error of the cylindrical radial coordinate~$r_{\rm cyl}$ with respect to the coordinate~\mbox{$r_{\rm Car}=\sqrt{x^2+y^2}$} in Cartesian geometry. As is evident, the error does not exceed~$2\times10^{-3}$ throughout the simulations.

\section{Drag}
\label{sect:drag}
\subsection{Implementation}
For our implementation of the drag exerted by the gas on the dust and vice versa, we take advantage of the first-order cloud-in-cell mapping between the grid and the particles that is part of the FLASH Code.

For each of the radial, azimuthal, and vertical velocity components, we execute the following steps:
\begin{enumerate}
  \item The gas density, pressure, and velocity component are mapped to the particles.
  \item For every particle:
  \begin{enumerate}
    \item The full-step dust velocity component is calculated from the stored half-step component according to Eq.~\ref{eq:full-step_velocity}.
    \item The dust stopping time~$t_{\rm d,stop}$ is computed as 
    \begin{equation}
    t_{\rm d,stop}=
    \begin{cases}
    \frac{\rho_{\rm s} a}{\rho_{\rm g} c_{\rm s}}&a\leq9/4\,\lambda_{\rm g,mfp}\text{ (Epstein regime)}\text{ or}\\
    \frac{4\rho_{\rm s} a^2}{9\rho_{\rm g} c_{\rm s}\lambda_{\rm g,mfp}}&a>9/4\,\lambda_{\rm g,mfp}\text{ (Stokes regime)},\\
    \end{cases}
    \end{equation}
    where~\mbox{$\lambda_{\rm g,mfp}=1/(\sigma_{\rm g} n_{\rm g})=\mu m_H/(\sigma_{\rm g} \rho_{\rm g})$} is the gas mean free path length,~$n_{\rm g}$ the gas number density,~\mbox{$\sigma_{\rm g}=2\times10^{-15}~\cm^2$} the molecular collision cross section \citep{Chapman1970}, and~\mbox{$c_{\rm s}=\gamma P/\rho_{\rm g}$} the sound speed.
     \item To the stored dust half-step velocity component, the drag source term~\mbox{$-\Delta v_{\rm d,drag}=-(v_{\rm d}-v_{\rm g})/t_{\rm d,stop}~\Delta t$} is added, where~$v_{\rm d}$ and~$v_{\rm g}$ are the (full-step) dust and gas velocity components, respectively, and~$\Delta t$ is the current time step.
  \end{enumerate}
  \item The change in dust momentum~\mbox{$\Delta p_{\rm d,drag}=m_{\rm d}\Delta v_{\rm d,drag}$} is mapped to the grid.
  \item The drag source term~\mbox{$\Delta v_{\rm g,drag}=\Delta p_{\rm d,drag}/(\rho_{\rm g} V)$} is added to the gas velocity in each grid cell, where~$V$ is the cell volume.
\end{enumerate}
We employ the global minimum of the particle stopping time and the gas stopping time~\mbox{$t_{\rm g,stop}=\rho_{\rm g}/\rho_{\rm d}~t_{\rm d,stop}$} as an upper limit of the simulation time step.

\subsection{Test}
To evaluate our algorithm, we adopt the test problem that is introduced by \citet{Laibe2011} and applied by \citet{Bai2010a}, \citet{Mignone2019} as well as \citet{Yang2016}, who refer to it as \textsc{dustybox}, particle-gas deceleration, and uniform streaming test, respectively. This problem provides an opportunity to simultaneously test the drag that is exerted by the gas on the dust and the drag back-reaction of the dust onto the gas. In addition, it permits us to compare the numeric solution which our algorithm computes to an analytic one.

We conduct simulations with a two-dimensional cylindrical geometry, as this is the geometry of the simulations we present in the main text. Dust and gas initially move in the radial direction with~\mbox{$v_{{\rm d},r,{\rm init}}=c_{\rm s}$} and~\mbox{$v_{{\rm g},r,{\rm init}}=-c_{\rm s}$}, respectively. The dust and gas density as well as the gas temperature are constant. Then, the equations of motion of the dust and the gas reduce to
\begin{equation}
\frac{{\rm d}v_{{\rm d},r}}{{\rm d}t}=-\frac{v_{{\rm d},r}-v_{{\rm g},r}}{t_{\rm d,stop}}\text{ and} 
\end{equation}
\begin{equation}
\frac{{\rm d}v_{{\rm g},r}}{{\rm d}t}=\epsilon\frac{v_{{\rm d},r}-v_{{\rm g},r}}{t_{\rm d,stop}},
\end{equation}  
where~$\epsilon$ is the solid-to-gas density ratio. The dust and the gas velocity can be solved for analytically, yielding
\begin{equation}
\begin{split}
v_{{\rm d},r}(t)=~&v_{{\rm d},r,{\rm init}}\exp\left[-(1+\epsilon)\frac{t}{t_{\rm d,stop}}\right]\\
&+v_{{\rm com},r,{\rm init}}\left(1-\exp\left[-(1+\epsilon)\frac{t}{t_{\rm d,stop}}\right]\right)\text{ and}
\end{split}
\end{equation}
\begin{equation}
\begin{split}
v_{{\rm g},r}(t)=~&v_{{\rm g},r,{\rm init}}\exp\left[-(1+\epsilon)\frac{t}{t_{\rm d,stop}}\right]\\
&+v_{{\rm com},r,{\rm init}}\left(1-\exp\left[-(1+\epsilon)\frac{t}{t_{\rm d,stop}}\right]\right),
\end{split}
\end{equation} 
where
\begin{equation}
v_{{\rm com},r,{\rm init}}=\frac{v_{{\rm g},r,{\rm init}}+\epsilon v_{{\rm p},r,{\rm init}}}{1+\epsilon}
\end{equation}
is the initial velocity of the center of mass of dust and gas.

The displacement of every dust particle relative to its initial position is given by
\begin{equation}
\begin{split}
\Delta r_{\rm d}(t)=~&r_{\rm d}(t)-r_{\rm d,init}\\
=~&\frac{(v_{{\rm d},r,{\rm init}}-v_{{\rm com},r,{\rm init}})\,t_{\rm d,stop}}{1+\epsilon}\left(1-\exp\left[-(1+\epsilon)\frac{t}{t_{\rm d,stop}}\right]\right)\\
&+v_{{\rm com},r,{\rm init}}\,t.
\end{split}
\label{eq:displacement}
\end{equation}
Initially, one particle is positioned at the center of every cell.

It is natural to choose the dust stopping time~$t_{\rm d,stop}$ and the sound speed~$c_{\rm s}$ as the units of time and velocity, respectively. Then, the unit of length is~$c_{\rm s}\,t_{\rm d,stop}$. The domains of our simulations span~$100~c_{\rm s}\,t_{\rm d,stop}$ in the radial dimension. The domain boundaries are periodic, which is necessary to maintain a constant dust and gas density and to conserve the total momentum of dust and gas. The simulations end after~$3~t_{\rm d,stop}$.  

\begin{figure*}[t]
\centering
\includegraphics[width=\textwidth]{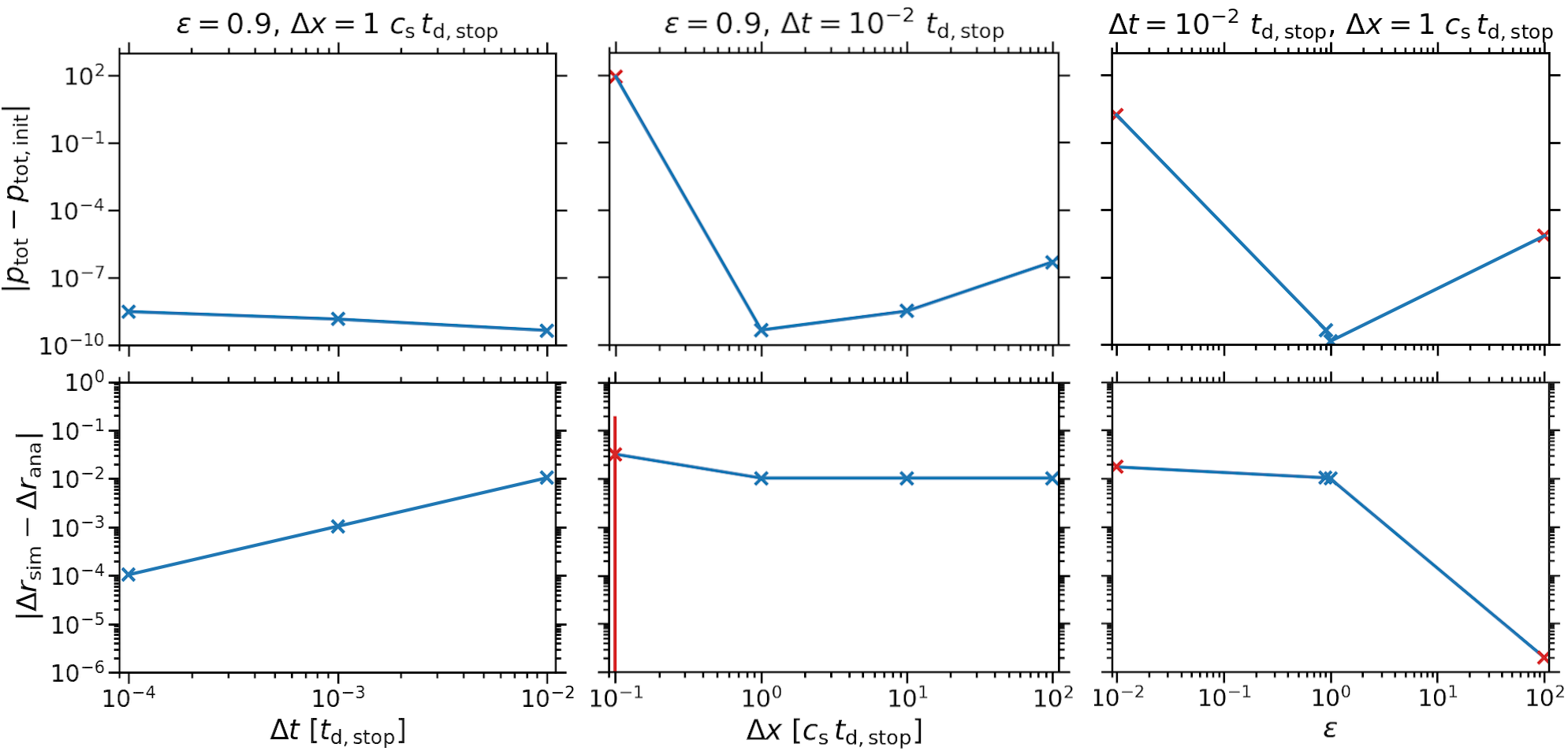}
\caption{Absolute error of the total momentum~$p_{\rm tot}$ of dust and gas relative to the initial total momentum~$p_{\rm tot, init}$ (upper panels) and mean absolute error of the simulated particle displacement~$\Delta r_{\rm sim}$ with respect to the analytic solution~$\Delta r_{\rm ana}$ (see Eq.~\ref{eq:displacement}; lower panels). In the latter case, the mean is calculated by averaging over all particles, with the standard deviations plotted as error bars. (With one exception, the standard deviations are too small for the error bars to be visible, though.) Both the momentum and the displacement errors are computed after~$3~t_{\rm d,stop}$. We show the errors as functions of the time step~$\Delta t$ (left panels), of the grid cell edge length~$\Delta x$ (middle panels), and of the dust-to-gas density ratio~$\epsilon$ (right panels). In the title of every column of panels, the fixed values of the respective other two quantities are given. Simulations in which particles cross the domain boundaries are marked with red crosses. The error in the total momentum can exceed one in these simulations, but is in general of the order of~$10^{-10}$ to~$10^{-9}$ otherwise. The error in the displacement increases linearly with the time step since our drag algorithm is first-order accurate. On the other hand, it is independent of the cell size for sizes of at least~$1~c_{\rm s}\,t_{\rm d,stop}$. At these resolutions, the particles are displaced by less than half a cell within~$3~t_{\rm d,stop}$ and thus do not transverse the boundaries. The error for the fiducial dust-to-gas density ratio of~$\epsilon=0.9$ amounts to~$10^{-2}$. While for a ratio of~$\epsilon=0.01$ it is greater by a factor of a few, it is as small as~${\sim}10^{-6}$ if the ratio is equal to~$\epsilon=100$. This is despite a number of particles crossing the boundaries if the density ratio is much less or much greater than the fiducial value.}
\label{fig:drag_test}
\end{figure*}

We employ two quantities to measure the accuracy of our implementation: the absolute error of the total momentum with respect to the initial value and the absolute error of the particle displacement relative to the analytic solution (see Eq.~\ref{eq:displacement}). In upper and lower panel of Fig.~\ref{fig:drag_test}, respectively, we show the momentum and the displacement error at the end of simulations with varying time steps~$\Delta t$ (left panels), grid cell edge lengths~$\Delta x$ (middle panels), and solid-to-gas density ratios~$\epsilon$ (right panels).

First and foremost, we note that our drag algorithm should not be used in combination with the periodic boundary implementation that is part of the FLASH Code. \footnote{This issue does not affect the simulations that are presented in the main text since we apply diode conditions at the boundaries of their domains.} In the figure, we mark with red crosses the errors in simulations in which particles cross the domain boundaries. The absolute error of the total momentum, while generally of the order of~$10^{-10}$ or~$10^{-9}$ otherwise, can be larger than unity in these simulations (see the upper panels).

We choose~$\epsilon=0.9$ as the fiducial dust-to-gas density ratio and~\mbox{$\Delta x=1~c_{\rm s}\,t_{\rm d,stop}$} as the fiducial resolution. This is to reduce the influence of the boundary conditions on our examination of the drag implementation as much as possible. For the fiducial density ratio, the analytic displacement remains less than~$0.436~c_{\rm s}\,t_{\rm d,stop}$, i.e., less than half a cell edge length at the fiducial resolution, within~$3~t_{\rm d,stop}$. Indeed, no particle crosses the boundaries in our simulations with this density ratio and the fiducial or a lower resolution.

From the lower left panel of the figure, it can be seen that our implementation is first-order accurate in time. For the fiducial time step of~\mbox{$\Delta t=10^{-2}$}, the absolute error of the displacement amounts to~$1\%$. If the time step is ten or a hundred times smaller, the error is less by one order or two orders of magnitude, respectively.

Reducing the resolution by a factor of ten or a hundred with respect to the fiducial resolution does not lead to an increase in the displacement error (see the lower middle panel). This is despite the displacement within~$3~t_{\rm d,stop}$ not being resolved even at the fiducial resolution. Nonetheless, the momentum error is greater at lower resolutions. It is of the order of~$10^{-7}$ in the simulation with a resolution of a hundredth of the fiducial resolution, in which only one particle is present. On the other hand, if the resolution is higher than the fiducial one, the error in both the displacement and the total momentum is considerable. This is because a large number of particles transverse the domain boundaries.

Compared to the error for the fiducial dust-to-gas density ratio, the displacement error increases by a factor of a few for low density ratios, but decreases by orders of magnitude for high ratios. This is evident from the lower right panel, in which the errors in simulations with the fiducial density ratio as well as~$\epsilon=0.01$, $\epsilon=1$, and~$\epsilon=100$ are shown. A number of particles crosses the boundaries in both the simulation with the highest density ratio and the one with the lowest ratio. Thus, the error in the total momentum, though not the error in the displacement, is significantly greater for these ratios than for the fiducial one.
\end{appendix}

\end{document}